\documentclass[12pt]{article}
\pdfoutput=1
\usepackage{putex}
\usepackage{comment}
\usepackage{graphicx}
\usepackage{caption}
\usepackage{amsmath}
\usepackage{array}
\usepackage{subcaption}
\usepackage{epstopdf}
\usepackage{enumerate}
\usepackage{cite}
\usepackage{youngtab}
\usepackage{tensor}
\usepackage{slashed}
\usepackage[export]{adjustbox}
\usepackage[aligntableaux=center]{ytableau}
\usepackage[utf8]{inputenc}
\usepackage[
      colorlinks=true,
      linkcolor=blue,
      urlcolor=blue,
      filecolor=black,
      citecolor=red,
      ]{hyperref}

\usepackage{braket}

\usepackage{tikz}
\usepackage{subcaption}

\newcommand{\RNum}[1]{\uppercase\expandafter{\romannumeral #1\relax}}

\newcommand {\be} {\begin {equation}}
\newcommand {\ee} {\end {equation}}

\newcommand {\bes} {\begin {equation*}}
\newcommand {\ees} {\end {equation*}}

\newcommand{\RP}{\mathbb{RP}}

\newcommand{\beq}{\begin{equation}}
\newcommand{\eeq}{\end{equation}}

\def\eqref#1{(\ref{#1})}

\def\ie{\begin{equation}\begin{aligned}}
\def\fe{\end{aligned}\end{equation}}

\numberwithin{equation}{section}

\def\<{\langle}
\def\>{\rangle}

\begin{document}

\preprint{PUPT-2618}

\institution{PU}{Department of Physics, Princeton University, Princeton, NJ 08544, USA}

\title{{\LARGE Wilson loops in $\mathcal{N}=4$ $SO(N)$ SYM \\ and D-Branes in $AdS_5\times \mathbb{RP}^5$}}

\authors{Simone Giombi and Bendeguz Offertaler}

\abstract{We study the half-BPS circular Wilson loop in ${\cal N}=4$ super Yang-Mills with orthogonal gauge group. By supersymmetric localization, its expectation value can be computed exactly from a matrix integral over the Lie algebra of $SO(N)$. We focus on the large $N$ limit and present some simple quantitative tests of the duality with type IIB string theory in $AdS_5\times \mathbb{RP}^5$. In particular, we show that the strong coupling limit of the expectation value of the Wilson loop in the spinor representation of the gauge group precisely matches the classical action of the dual string theory object, which is expected to be a D5-brane wrapping a $\mathbb{RP}^4$ subspace of $\mathbb{RP}^5$. We also briefly discuss the large $N$, large $\lambda$ limits of the $SO(N)$ Wilson loop in the symmetric/antisymmetric representations and their D3/D5-brane duals. 
Finally, we use the D5-brane description to extract the leading strong coupling behavior of the ``bremsstrahlung function" associated to a spinor probe charge, or equivalently the normalization of the two-point function of the displacement operator on the spinor Wilson loop, and obtain agreement with the localization prediction.} 

\maketitle

\tableofcontents

\section{Introduction}
\label{sec:intro}

In addition to being fundamental observables in gauge theory, Wilson loop operators are also versatile probes of the AdS/CFT correspondence  \cite{maldacena1999large,gubser1998gauge,witten1998anti}. They provide a rather direct manifestation of the gauge/string duality: a Wilson loop in the fundamental representation of the gauge group is dual to a fundamental open string worldsheet ending at the boundary of AdS along the curve defining the loop operator \cite{maldacena1998wilson, Rey:1998ik}. In the ${\cal N}=4$ SYM theory, it is natural to consider a generalization of the usual Wilson loop, sometimes referred to as the ``Maldacena-Wilson" loop, by including couplings to the adjoint scalar fields of the theory.\footnote{On the dual AdS side, these couplings are reflected in the Dirichlet boundary conditions for the string surface along the directions of the ``internal" space corresponding to the scalar fields, i.e. $S^5$ or $\RP^5$ for ${\cal N}=4$ SYM. The ordinary Wilson loop with no scalar couplings corresponds instead to Neumann boundary conditions on the internal space directions \cite{Alday:2007he, Polchinski:2011im, Beccaria:2017rbe, Beccaria:2019dws}. In this paper, we will only consider the Maldacena-Wilson loop, and will often refer to it simply as the Wilson loop.} This allows in particular the construction of loop operators that preserve various fractions of the superconformal symmetry of the theory \cite{zarembo2002supersymmetric,Drukker:2007qr}. The most supersymmetric example is the half-BPS Wilson loop, which is defined on a circular (or straight line) contour and couples to one of the six scalars of SYM theory. Its expectation value for circular contours is a non-trivial function of the coupling and representation of the gauge group that can be computed analytically by supersymmetric localization \cite{erickson2000wilson, drukker42exact, pestun2007localization}. Such exact results can be used to conduct non-trivial tests of the AdS/CFT duality and of the detailed structure of the gauge/string dictionary. In particular, one can make contact with the dynamics of various stringy objects in the dual theory: while Wilson loops in the fundamental representation are described by string worldsheets, large representations with ``size"\footnote{By ``size" we mean essentially $\sim \sum_i m_i$, where the $m_i$'s are the Young tableaux labels specifying the representation.} of order $N$ are dual to D-branes that pinch on the boundary contour \cite{drukker2005all, yamaguchi2006wilson, gomis2006holographic}, and even larger representations of order $N^2$ are dual to new supergravity backgrounds \cite{Yamaguchi:2006te, Lunin:2006xr, DHoker:2007mci}. 

Most of the existing work in the literature focuses on the duality between ${\cal N}=4$ SYM with $SU(N)$ gauge group and type IIB string theory on $AdS_5 \times S^5$. In this note, our aim is to provide some new quantitative tests of the duality in the case of ${\cal N}=4$ SYM with orthogonal gauge group. The relevant string theory dual was identified in \cite{witten1998baryons}. In flat space, one can obtain a $SO(N)$ or $USp(N)$ gauge theory by placing a stack of D3-branes at an orientifold 3-plane. In the near-horizon limit, one gets string theory on the $AdS_5\times \RP^5$ orientifold. Here $\RP^5=S^5/\mathbb{Z}_2$, where the $\mathbb{Z}_2$ action identifies antipodal points on $S^5$, yielding the five-dimensional real projective space. The difference between the $SO(N)$ and $USp(N)$ gauge theories lies in a choice of ``discrete torsion" for the Neveu-Schwarz and Ramond-Ramond $B$-fields \cite{witten1998baryons}. While the theories on $AdS_5\times S^5$ and $AdS_5\times \RP^5$ share several similarities, interesting new features arise in the latter. For example, the  genus expansion in string theory includes non-orientable worldsheets, which corresponds to the well-known fact that in $SO(N)$ and $USp(N)$ gauge theories the $1/N$ expansion includes both even and odd powers. Also, the possibility of wrapping branes on the internal $\RP^5$ gives rise to new objects that do not have counterparts in the $AdS_5\times S^5$/$SU(N)$ duality. For instance, in the $SO(2k)$ gauge theory one can construct ``Pfaffian" local operators ${\rm Pf}(\Phi)=\epsilon^{a_1\cdots a_{2k}}\Phi_{a_1a_2}\cdots \Phi_{a_{2k-1}a_{2k}}$ (where each $\Phi_{ab}$ is any of the six adjoint scalars of the theory, and we are being schematic by omitting the $R$-symmetry indices\footnote{To obtain a chiral primary operator, one should project onto the symmetric traceless representation of $SO(6)$.}), which are dual to D3-branes wrapping a $\RP^3\subset \RP^5$ \cite{witten1998baryons, Aharony:2002nd}. Another peculiarity of gauge theory with orthogonal gauge group, which is more relevant for the present paper, is that one can consider Wilson loops in the {\it spinor representation} of the gauge group. This is the familiar representation of $SO(N)$ whose generators are the rank-2 antisymmetric products of the Dirac gamma matrices (it has dimension $2^{N/2}$ for $N$ even\footnote{More precisely, for $N$ even it splits into two irreducible representations of opposite chirality.} and $2^{(N-1)/2}$ for $N$ odd). It does not have a counterpart in the $SU(N)$ or $USp(N)$ theories. As argued in \cite{witten1998baryons}, the half-BPS Wilson loop in the spinor representation is expected to be dual to a D5-brane wrapping a $\RP^4\subset \RP^5$ and occupying an $AdS_2$ subspace of $AdS_5$. In terms of Young tableaux labels, the spinor representation corresponds to $\vec{m}=[1/2,1/2,\ldots, 1/2]$, and hence $\sum_i m_i\sim N$. Thus, as in the case of large rank (anti)symmetric representations studied in \cite{drukker2005all, yamaguchi2006wilson, gomis2006holographic}, it is natural that the dual object should be a D-brane. The fact that Wilson loops in the spinor representation should be dual to a stringy object with tension of order $N$ (like a D-brane) was also foreshadowed by earlier work \cite{Witten:1983tx}. 

In this paper, we use supersymmetric localization to obtain a simple quantitative test of the correspondence between the spinor Wilson loop and the $AdS_2\times \RP^4$ D5-brane. Although this gauge theory observable has already been studied at finite $N$ and 't Hooft coupling $\lambda=g_{YM}^2N$ \cite{fiol2014exact}, here we focus on the large $N$, large $\lambda$ regime and provide a direct confirmation that the expectation value in the gauge theory matches the classical action of the wrapped D5-brane of \cite{witten1998baryons}. We also consider the fundamental Wilson loop (mainly as a check of numerical factors in the gauge/string dictionary and of the consistency of our conventions for the gauge theory and matrix integral), and Wilson loops in the large (anti)symmetric representations. The latter are expected to be dual to D3 and D5-branes with $AdS_2\times S^2$ and $AdS_2\times S^4$ worldvolumes, similarly to the previously studied examples \cite{drukker2005all, yamaguchi2006wilson, gomis2006holographic} in the $AdS_5\times S^5$/$SU(N)$ case. 

The study of the half-BPS Wilson loop in ${\cal N}=4$ SYM and its deformations has recently received renewed attention due to its connection to defect CFT. This is because the circular (or straight) half-BPS loop preserves a $SL(2,\mathbb{R})$ 1d conformal symmetry and hence can be viewed as a conformal defect \cite{drukker2006small, Cooke:2017qgm, giombi2017half, liendo2018bootstrapping} labelled by the choice of representation for the probe particle running around the loop. The case of the fundamental representation of $SU(N)$ has been studied extensively both on the gauge theory and string theory sides (see e.g. \cite{Cooke:2017qgm, giombi2017half, Kim:2017sju, Kiryu:2018phb, giombi2018exact}). Likewise, defect CFT correlators on Wilson loops in (anti)symmetric representations and their duality to fluctuations of the D3 and D5-branes were recently studied in detail in \cite{Giombi:2020amn}. It would be interesting to perform similar defect CFT analyses for $SO(N)$ gauge theory and in particular for the spinor representation, which is a unique feature of orthogonal groups. In this paper, we take a first step towards such a study: we will only focus on the transverse fluctuations of the D5-brane within $AdS_5$, which are dual to the displacement operators of the defect CFT. The normalization of their two-point function is given by the so-called ``bremsstrahlung function" \cite{correa2012exact}, which can be computed exactly for any representation using supersymmetric localization. Using the D5-brane and applying the $AdS_2/dCFT_1$ correspondence along the lines of \cite{giombi2017half}, we obtain the leading strong coupling prediction for the spinor bremsstrahlung function, finding agreement with the localization result. We leave a more detailed analysis of the D5-brane fluctuations and their defect CFT dual for future work. 

The rest of this paper is organized as follows. Section~\ref{sec:WgLkoZOJwt} focuses on the calculation of the circular Wilson loop expectation value in the ${\cal N}=4$ $SO(N)$ SYM using supersymmetric localization, which reduces it to a matrix integral over the Lie algebra of $SO(N)$. In \S~\ref{sec:gghp3J2AIM}, we compute the expectation value of the Wilson loop in the fundamental representation using the method of orthogonal polynomials, mainly as a check of our matrix model conventions. 
%and compare the leading terms at small coupling to perturbation theory; this ensures that our conventions for the matrix integral match our conventions for the Wilson loop and $\mathcal{N}=4$ SYM action. 
In \S~\ref{sec:J6TEzJQSXU}, we derive the Wigner semicircle law characterizing the distribution of the eigenvalues at the saddle point of the matrix integral at large $N$, and use it to derive the large $N$ limit of the Wilson loop in the spinor representation in \S~\ref{sec:E4dda6KwCr} and in the large rank (anti)symmetric representations in \S~\ref{sec:cAhVgwXzVQ}. In \S~\ref{sec:jKYRKLXLoH}, we focus on the dual string theory description. After a brief review of the $AdS_5\times \RP^5$ dual and of the matching between fundamental representation and fundamental string, we show in \S~\ref{sec:HReGcbyOsR} that the value of the classical action of the $AdS_2\times \RP^4$ D5-brane precisely agrees with the gauge theory calculation in \S~\ref{sec:E4dda6KwCr}. In \S~\ref{sec:0pfY1VVBbg}, we describe the D3/D5-branes dual to the Wilson loop in the large rank symmetric and antisymmetric representations. Section~\ref{sec:MOwof1jOHQ} is devoted to the two point function of the displacement operators in the 1d defect CFT defined by the spinor Wilson loop. We calculate the two-point function using both supersymmetric localization and the $AdS_2$ theory of fluctuations of the wrapped D5-brane, and find agreement. Finally, in \S~\ref{sec:ewr5bkUoCk}, we summarize our results and suggest possible extensions to our work.

\section{Wilson loops in $\mathcal{N}=4$ $SO(N)$ SYM}\label{sec:WgLkoZOJwt}
We begin this section by introducing the $SO(N)$ half-BPS Maldacena-Wilson loop and the matrix model to which it is equivalent via supersymmetric localization. We then apply the saddle point method to the matrix integral to determine the expectation value of the large $N$ Wilson loop in various representations, which we will compare to the classical actions of the corresponding strings and D-branes in $AdS_5\times \RP^5$ in \S~\ref{sec:jKYRKLXLoH}.

First, we introduce our conventions. We normalize the gauge and scalar fields of $\mathcal{N}=4$ SYM so that their kinetic terms take the following standard form:
\begin{align}\label{eq:yhbwoLehbV}
    S_{SYM}&=\frac{1}{g_{YM}^2}\int d^4x \text{tr}\left\{\frac{1}{2}F_{\mu\nu}F^{\mu\nu}+(\partial_\mu \Phi^I)^2+\ldots\right\}.
\end{align}
Here $F_{\mu\nu}=\partial_{[\mu}A_{\nu]}-i[A_\mu,A_\nu]$ is the gauge field strength, $A_\mu=A_\mu^a T^a$ is the gauge field, $\Phi^I=\Phi^I_aT^a$ are the six scalar fields, and $T^a$ are the generators of the Lie algebra $\mathfrak{g}$ satisfying $\text{Tr}(T^aT^b)=C(F)\frac{\delta^{ab}}{2}$ (the Lie algebra indices $a,b$ run from $1$ to $\text{dim}(\mathfrak{g})$). Here $C(F)$ corresponds to a choice of normalization for the generators in the fundamental representation. A conventional value is $C(F)=1/2$, but we will keep it arbitrary below (nothing will depend on the choice of $C(F)$). Both the gauge field and the scalar fields transform in the adjoint representation of the gauge group, $G$. 

The Maldacena-Wilson loop is defined to be the following path-ordered exponential %\cite{maldacena1998wilson,drukker42exact,erickson2000wilson,drukker1999wilson,zarembo2002supersymmetric}
\begin{align}\label{eq:2W2phlg0Lb}
    W_R&=\text{Tr}_R \mathcal{P}\left[\oint \left(iA_\mu (x) \dot{x}^\mu+ \Theta_I\Phi^I(x) |\dot{x}|\right)dt\right],
\end{align}
where $x^\mu(t)$ is a closed path in $\mathbb{R}^{4}$, and $\Theta^I(t)$ is a unit 6-vector (i.e., $\Theta^I\Theta^I=1$), 
%\footnote{The coupling $\Theta_I \Phi^I$ distinguishes the Maldacena-Wilson loop from the usual Wilson loop discussed in Yang-Mills theories, and corresponds to fixing Dirichlet boundary conditions on the compact submanifold $X$ of $AdS_5\times X$.} 
and $R$ is a choice of representation characterizing the external probe particle propagating along the contour. Our primary focus will be on the spinor representation of $G=SO(N)$, but we will also discuss the fundamental, rank $k$ symmetric and rank $k$ antisymmetric representations.

We restrict our attention to the half-BPS Wilson loops, for which $x^\mu(t)$ is a circle and $\Theta^I(t)$ is a constant, corresponding to choosing one of the six scalars. The operator in Eq.~\eqref{eq:2W2phlg0Lb} then preserves half of the supersymmetry.\footnote{The straight Wilson line is also half-BPS, but its expectation value is equal to the representation dimension for any representation and coupling.} By supersymmetric localization \cite{pestun2007localization}, the expectation value of the half-BPS Wilson loop for a given gauge group $G$ reduces to an integral over the elements of the Lie algebra $\mathfrak{g}$ of $G$,
\begin{align}\label{eq:g8ug8suaiq}
    \braket{W_R}&=\frac{1}{Z}\int_{\mathfrak{g}} \mathcal{D}K \text{Tr}_R(e^{K})\text{exp}\left(-\frac{2}{g_{YM}^2}\text{Tr}K^2\right),
\end{align}
where $K=K^a T^a$ are Lie algebra elements, and $Z=\int \mathcal{D}K\text{exp}\left(-\frac{2}{g_{YM}^2}\text{Tr}K^2\right)$ is the partition function of the matrix model. We can check that the normalization of the action in \eqref{eq:g8ug8suaiq} agrees with the gauge theory conventions in Eq.~\eqref{eq:yhbwoLehbV} by computing $\braket{W_R}$ perturbatively. Noting that $\text{Tr}K^2 = K^a K^b \text{Tr}(T^a T^b)=C(F) K^a K^a$, so that the propagator is $\langle K^a K^b\rangle = \frac{g_{YM}^2}{4C(F)}\delta^{ab}$, one finds 
\begin{equation}
\langle W_R\rangle = \text{dim}(R)+\frac{g_{YM}^2}{8C(F)}C_2(R)\text{dim}(R)+\ldots 
\label{MMpert}
\end{equation} 
where the quadratic Casimir is defined by $T^a_RT^a_R=C_2(R)1_{\text{dim}(R)\times \text{dim}(R)}$. To compare with the perturbative expansion of the Wilson loop in the SYM theory, we expand the exponential in Eq.~\eqref{eq:2W2phlg0Lb} and use the gauge and scalar field propagators. The latter follow from the action, Eq.~\eqref{eq:yhbwoLehbV}, which in the Feynman gauge gives:
\begin{equation}
\braket{A_\mu^a(x) A_\nu^b(y)}=\frac{g_{YM}^2}{8\pi^2 C(F)}\frac{\delta_{\mu\nu}\delta^{ab}}{|x-y|^2}\,,\qquad  \braket{\Phi^{Ia}(x)\Phi^{Jb}(y)}
=\frac{g_{YM}^2}{8\pi^2C(F)}\frac{\delta^{ab}\delta^{IJ}}{|x-y|^2}\,.
\end{equation}
For the circular contour $x^\mu(t)=(\cos{t},\sin{t},0,0)$ the ``combined" gauge field and scalar propagator is a constant 
\begin{align}
    -\braket{A_\mu^a(x(t))A_\nu^b(x(s))}\dot{x}^\mu(t) \dot{x}^\nu(s)+\braket{\Theta_I\Phi^{Ia}(x(t))\Theta_J\Phi^{Jb}(x(s))}|\dot{x}(t)||\dot{x}(s)|
=\frac{g_{YM}^2\delta^{ab}}{16\pi^2 C(F)}
\end{align}
and one finds
\begin{align}
    \braket{W_R}&=\text{dim}(R)+\frac{1}{2}\int_{0}^{2\pi}dt \int_0^{2\pi}ds \text{Tr}_R\{\braket{(iA_\mu\dot{x}^\mu +\Theta_I\Phi^I|\dot{x}|)^2}\}+O(g_{YM}^3)\nonumber\\
&=\text{dim}(R)+\frac{g_{YM}^2}{8C(F)}C_2(R)\text{dim}(R)+\ldots 
\end{align}
in agreement with Eq.~\eqref{MMpert}. 

The discussion so far applies to arbitrary gauge group. Let us now specialize to the case of $SO(N)$, which is the main focus of this paper. In this case, the Lie algebra consists of the antisymmetric Hermitian $N\times N$ matrices $K_{jk}=K_{kj}^*=-K_{kj}$ and the measure in the matrix model may be written as $\mathcal{D}K=\prod_{j<k}d(-iK_{jk})$. Below, we review some of the relevant tools to analyze the large $N$ limit of the matrix model (see e.g. \cite{brezin1978,bessis1980quantum,marino2004houches} for general discussions of random matrix integral techniques). We will use $\lambda =g_{YM}^2 N$ to denote the 't Hooft coupling in the $SO(N)$ gauge theory, which is held fixed as $N\rightarrow \infty$.\footnote{An alternative definition of $\lambda$ is to let it be $g_{YM}^2$ times the rank of $SO(N)$. In that case, $\lambda=g_{YM}^2\frac{N}{2}$ for $N$ even and $\lambda=g_{YM}^2\frac{N-1}{2}$ for $N$ odd. We will not adopt this definition in this paper.}   

Antisymmetric Hermitian matrices can be block-diagonalized using orthogonal matrices \cite{zumino1962normal}. Since the traces and the measure in Eq.~\eqref{eq:g8ug8suaiq} are invariant under $K\to OKO^T$ if $OO^T=I$, conjugation by orthogonal matrices is a gauge symmetry of Eq.~\eqref{eq:g8ug8suaiq} that can be fixed using the Faddeev-Popov procedure. The resulting Faddeev-Popov determinant, denoted $\Delta^2(\Gamma)$, can alternatively be viewed as the Jacobian for a change of variables from $K$ to $O\Gamma O^T$, $O$ being orthogonal and $\Gamma$ being block-diagonal. The resulting integral over the eigenvalues $\gamma_i$ is slightly different for the even ($N=2N'$) and odd ($N=2N'+1$) cases. One finds (see e.g. \cite{mehta1968distribution, mehta1983some, fiol2014exact}):
\begin{align}\label{eq:25ItjrOuK2}
\braket{W_R}&=\frac{1}{Z'}\int\left(\prod_{i=1}^{N'}d\gamma_i\right)\Delta^2(\Gamma)\text{Tr}_R(e^{\Gamma})\text{exp}\left(-\frac{2N}{\lambda}\text{Tr}\Gamma^2\right),
\end{align}
where
\begin{align}\label{eq:VcM3skO8Vn}
    \Gamma=\text{diag}\left(\gamma_1\sigma_y,\gamma_2\sigma_y,\ldots,\gamma_{N'}\sigma_y\right),\hspace{1cm}\Delta^2(\Gamma)=\prod_{1\leq j<k\leq N'} |\gamma_j^2-\gamma_k^2|^2
\end{align}
if $N=2N'$, and where
\begin{align}\label{eq:tj14e0SXd7}
     \Gamma=\text{diag}\left(\gamma_1\sigma_y,\gamma_2\sigma_y,\ldots,\gamma_{N'}\sigma_y,0\right),\hspace{1cm}\Delta^2(\Gamma)=\prod_{1\leq j<k\leq N'} |\gamma_j^2-\gamma_k^2|^2\prod_{l=1}^{N'}\gamma_l^2
\end{align}
if $N=2N'+1$. In Eqs.~(\ref{eq:VcM3skO8Vn}) and (\ref{eq:tj14e0SXd7}), $\sigma_y=\begin{pmatrix}0&-i\\i&0\end{pmatrix}$ is the second Pauli matrix. In Eq.~\eqref{eq:25ItjrOuK2} the normalization factor $Z'$ is set equal to the integral expression in Eq.~\eqref{eq:25ItjrOuK2} with $\text{Tr}_R(e^\Gamma)$ replaced by $1$. This ensures that the expectation value of the Wilson loop is equal to the dimension of the representation in the free limit (i.e., $\braket{W_R}\rvert_{\lambda\to 0}=\text{dim }R$). For later use, we also note that $\text{Tr} \Gamma^2=2\sum_{i=1}^{N'}\gamma_i^2$ and the eigenvalues of $e^\Gamma$ are $\{e^{\pm \gamma_i}|i=1,\ldots,N'\}$ for $N=2N'$ and $\{1\}\cup \{e^{\pm \gamma_i}|i=1,\ldots,N'\}$ for $N=2N'+1$. 

One can in principle use the method of orthogonal polynomials to compute Eq.~\eqref{eq:25ItjrOuK2} exactly for any representation $R$ and finite $N$. The product $\prod_{i<j}(\gamma_i^2-\gamma_j^2)$ appearing in $\Delta(\Gamma)$ is the determinant of the Vandermonde matrix with entries $V_{ij}=(\gamma_i^2)^{j-1}$. Using the column addition identity of the determinant, one can write $|V_{ij}|=|P_j(\gamma_i)|$, where $P_n(x)$ is any even monomial in $x$ of order $2(n-1)$. The integral in Eq.~\eqref{eq:25ItjrOuK2} simplifies if we pick the polynomials $P_n(x)$ to be orthogonal when integrated with the weight $e^{-\frac{4N}{\lambda}x^2}$ for $N=2N'$ and the weight $e^{-\frac{4N}{\lambda}x^2}x^2$ for $N=2N'+1$. The orthogonality helps separate the integrals over the different eigenvalues. The right choices for the even and odd cases are $P_n(x)\propto H_{2n-2}(2\sqrt{N/\lambda}x)$ and $P_n(x)\propto \frac{1}{x}H_{2n-1}(2\sqrt{N/\lambda}x)$, respectively, where $H_n(x)$ is the $n$th Hermite polynomial. More details can be found in \cite{fiol2014exact}. 

In this paper we are mainly interested in the large $N$ limit, and to analyze higher representations in this limit it will be more convenient to use the saddle point method and corresponding density of eigenvalues. Before doing so, in the next section we will briefly discuss the evaluation of the Wilson loop in the fundamental representation using the orthogonal polynomial method. This will serve as a consistency check of the large $N$ eigenvalue distribution  we will derive in $\S$~\ref{sec:J6TEzJQSXU}.

\subsection{Fundamental representation\label{sec:gghp3J2AIM}}

We now apply the method of orthogonal polynomials to determine the expectation value of the Wilson loop in the fundamental representation, which we denote $R=F$. In this representation, $\text{Tr}(e^\Gamma)=2\sum_{i=1}^{N'}\cosh(\gamma_i)$ and $\text{Tr}(e^\Gamma)=1+2\sum_{i=1}^{N'}\cosh(\gamma_i)$ in the even and odd cases, respectively. Because $\text{Tr}(e^\Gamma)$ in Eq.~\eqref{eq:25ItjrOuK2} depends additively on the $\cosh(\gamma_i)$, the Wilson loop expectation value reduces to a sum of single integrals involving Hermite polynomials that can be evaluated in terms of associated Laguerre polynomials. Specifically:
\begin{align}\label{eq:FQtTqkvJ3O}
    \braket{W_F}&=\frac{2}{\sqrt{\pi}}\sum_{i=0}^{N'-1}\frac{1}{(2i)!2^{2i}}\int dx H_{2i}^2(x)\cosh\left(\sqrt{\frac{\lambda}{N}}\frac{x}{2}\right)e^{-x^2}\nonumber\\&=2e^{\frac{\lambda}{16N}}\sum_{i=0}^{N'-1}L_{2i}\left(-\frac{\lambda}{8N}\right)
\end{align}
for $N=2N'$, and
\begin{align}\label{eq:zNZmwZWmi1}
   \braket{W_F}&=1+\frac{2}{\sqrt{\pi}}\sum_{i=0}^{N'-1}\frac{1}{(2i+1)!2^{2i+1}}\int dx H_{2i+1}^2(x)\cosh\left(\sqrt{\frac{\lambda}{N}}\frac{x}{2}\right)e^{-x^2}\nonumber\\&=1+2e^{\frac{\lambda}{16N}}\sum_{i=0}^{N'-1}L_{2i+1}\left(-\frac{\lambda}{8N}\right)
\end{align}
for $N=2N'+1$. These expressions agree with those in \cite{fiol2014exact}. Using a contour integral representation of the Laguerre polynomials,\footnote{One may use $L_n(x)=(2\pi i)^{-1}\oint dt (t+x)^n t^{-n-1}e^{-t}$, and expand the sum of Laguerre polynomials as a series in $1/N$. This leads to contour integrals of the form $(2\pi i)^{-1}\oint du e^{-\sqrt{\lambda/8}(u+u^{-1})}u^{-n}$, which give Bessel functions.}
%Finding the residues of some of the resulting terms--- those of the general form $e^{-\lambda/(8t)}e^{-t}r(t)$, where $r(t)$ is a rational function and $e^{-\lambda/(8t)}$ has an essential singularity at $t=0$--- means reducing them to contour integrals of the form $(2\pi i)^{-1}\oint du e^{-\sqrt{\lambda/8}(u+u^{-1})}u^{-n}$, which give Bessel functions.} 
it is not difficult to determine the first few terms in the $1/N$ expansion (with $\lambda$ fixed) of the circular Wilson loop expectation value directly from Eqs.~(\ref{eq:FQtTqkvJ3O}) and (\ref{eq:zNZmwZWmi1}). For both the even and odd cases,
\begin{align}\label{eq:1MQjpdsJek}
\frac{1}{N}\braket{W_F}&=\frac{2\sqrt{2}}{\sqrt{\lambda}}I_1\left(\sqrt{\frac{\lambda}{2}}\right)+\frac{1-I_0\left(\sqrt{\frac{\lambda}{2}}\right)}{2N}+\frac{\lambda I_2\left(\sqrt{\frac{\lambda}{2}}\right)}{96N^2}+O(1/N^3).
\end{align}
We can compare the small $g_{YM}$ expansion of Eq.~\eqref{eq:1MQjpdsJek}, %$\frac{1}{N}\braket{W_F}=1+\frac{g_{YM}^2}{16}\left(N-1\right)+O(g_{YM}^4)$, 
with the perturbative expansion in Eq.~(\ref{MMpert}). 
When $R=F$ is the fundamental representation of $G=SO(N)$ with $\text{dim}(R=F)=N$, the quadratic Casimir is $C_2(F)=C(F)(N-1)/2$ (see the Appendix), so that (\ref{MMpert}) yields $N^{-1}\braket{W_F}=1+g_{YM}^2(N-1)/16+O(g_{YM}^4)$. This agrees with the small $g_{YM}$ expansion of Eq.~\eqref{eq:1MQjpdsJek}. 
%Given our conventions for the SYM action, stated in Eq.~\eqref{eq:yhbwoLehbV}, and for the Wilson loop, stated in Eq.~\eqref{eq:2W2phlg0Lb}, the agreement confirms that the normalization of the exponent in Eq.~\eqref{eq:g8ug8suaiq} is correct.

\subsection{Large $N$ saddle point of the $SO(N)$ matrix model}\label{sec:J6TEzJQSXU}

Computing the large $N$ behavior of the Wilson loop, Eq.~\eqref{eq:25ItjrOuK2}, via the orthogonal polynomial method is more cumbersome for higher representations. In the present section, we use the saddle point method and obtain the density of eigenvalues (given essentially by Wigner's semicircle law) for the $SO(N)$ matrix integral. Our analysis is similar to the one in \cite{ashok2003unoriented}.

Consider the large $N'$ regime of the following $N'$-dimensional integral
\begin{align}
    \braket{f}&=\frac{1}{Z'}\int \left(\prod_{i=1}^{N'}d\gamma_i\right)f(\gamma_1,\ldots,\gamma_{N'})e^{-N'^2U(\gamma_1,\ldots,\gamma_{N'})},\label{eq:P2iNDsfwDy}
\end{align}
where the normalization $Z'$ is chosen so that $\braket{1}=1$ and $U=-\frac{1}{{N'}^2}\log\left(\Delta^2(\Gamma)e^{-\frac{2N}{\lambda}\text{Tr}\Gamma^2}\right)$ is the logarithm of the measure of Eq.~\eqref{eq:25ItjrOuK2}. Let us start with the 
even $N$ case, i.e. $N=2N'$. Then we have
\begin{align}
    U(\gamma_1,\ldots,\gamma_{N'})&=\frac{8}{\lambda N'}\sum_{i=1}^{N'}\gamma_i^2-\frac{2}{N'^2}\sum_{1\leq j<k\leq N'}\log|\gamma_j^2-\gamma_k^2|.\label{eq:hoe7tK35y6}
\end{align}
%In order to reasonably interpret the large $N'$ limit of $\braket{f}$, the function $f(\gamma_1,\ldots,\gamma_{N'})$ should be the $N'$th member of a class of functions that are indexed by the number of independent variables of which they are functions. For simplicity, 
We can restrict our attention to functions of the form $f(\gamma_1,\ldots,\gamma_{N'})=g(\sum_{i=1}^{N'} h(\gamma_i))$, which is general enough for Eq.~\eqref{eq:P2iNDsfwDy} to encompass Eq.~\eqref{eq:25ItjrOuK2} as a special case. 

Eq.~\eqref{eq:P2iNDsfwDy} is dominated in the $N'\to \infty$ limit by the eigenvalue configurations that minimize the ``potential'' $U(\gamma_1,\ldots,\gamma_{N'})$. The invariance of the potential under $\gamma_i\to -\gamma_i$ for any $i=1,2,\ldots,N'$ means that there are $2^{N'}$ distinct configurations (or $2^{N'-1}$ if one of the eigenvalues is zero) that minimize $U$. (We do not consider configurations related by permutations of the eigenvalues to be distinct). The degeneracy of the minimum means the minimizing eigenvalues are not specified by a single distribution. By contrast, the distribution of the squares of the eigenvalues, which are also invariant under $\gamma_i\to -\gamma_i$, is the same for all $\sim 2^{N'}$ configurations. This unique distribution approaches a smooth function, $\sigma (\gamma^2)$, in the large $N'$ limit.

Our task is to find $\sigma(\gamma^2)$. The eigenvalues minimizing $U$ satisfy $\partial U/\partial \gamma_i=0$, which yields:
\begin{align}\label{eq:cHnPl8iYIA}
    \frac{4\gamma_i}{\lambda}=\frac{\gamma_i}{N'}\sum_{j\neq i}\frac{1}{\gamma_i^2-\gamma_j^2},\hspace{1cm}\text{for }i=1,\ldots,N'.
\end{align}
One must not cancel $\gamma_i$ from both sides because one of the eigenvalues may be zero (see footnote \ref{ftnt:1}). Instead, we multiply both sides of Eq.~\eqref{eq:cHnPl8iYIA} by $\gamma_i/(z-\gamma_i^2)/z$ and sum over $i$. Writing  $\gamma_i^2/(z-\gamma_i^2)/z=-1/z+1/(z-\gamma_i^2)$, the left hand side yields two terms and, since $\sum_i \sum_{j\neq i}(\gamma_i^2-\gamma_j^2)^{-1}=0$, the right hand side simplifies. It follows that:
\begin{align}\label{eq:SuS7Svx0qL}
    \frac{4}{\lambda}\left(-\frac{1}{z}+G(z)\right)&=\frac{1}{{N'}^2}\sum_i \frac{1}{z-\gamma_i^2}\sum_{j\neq i}\frac{1}{\gamma_i^2-\gamma_j^2},
\end{align}
where we have introduced the resolvent, $G(z)$:
\begin{align}\label{eq:u56ww2j0Wa}
    G(z)=\frac{1}{N'}\sum_{i=1}^{N'}\frac{1}{z-\gamma_i^2}.
\end{align}
The resolvent is analytic on $\mathbb{C}\setminus[0,a]$, where $a\propto \lambda$ is the supremum of the squares of the eigenvalues in the large $N'$ limit. It satisfies $\text{Tr}(x-\Gamma)^{-1}=2xN' G(x^2)$, which  essentially makes $\braket{G(z)}$ a generating function for $\braket{\text{Tr}(\Gamma^n)}$. Assuming the squares of the eigenvalues at the minimum are characterized by a smooth distribution $\sigma(\gamma^2)$ in the large $N'$ limit, $G(z)$ may be written:
\begin{align}
    G(z)=\int_0^a \frac{\sigma(u)du}{z-u}.
\end{align}
The residue theorem then implies 
\begin{align}\label{eq:2GKGn244FH}
    \sigma(u)=\frac{1}{2\pi i}\left[G(u+i\epsilon)-G(u-i\epsilon)\right],
\end{align}
which means $G(z)$ determines $\sigma(u)$ and vice versa. 

To find $\sigma(u)$,  we derive a quadratic equation for $G(z)$ using Eq.~\eqref{eq:SuS7Svx0qL}. The square of the resolvent can be written:
\begin{align}
    G^2(z)&=\frac{1}{N'^2}\sum_{i=1}^{N'}\frac{1}{(z-\gamma_i^2)^2}+\frac{1}{N'^2}\sum_{i=1}^{N'}\sum_{j=1,j\neq i}^{N'}\frac{1}{z-\gamma_i^2}\frac{1}{z-\gamma_j^2}\nonumber\\&=-\frac{1}{N'}G'(z)+\frac{2}{N'^2}\sum_{i=1}^{N'}\frac{1}{z-\gamma_i^2}\sum_{j=1,j\neq i}^{N'}\frac{1}{\gamma_i^2-\gamma_j^2}.\label{eq:GW5CYUB2bA}
\end{align}
Applying Eq.~\eqref{eq:SuS7Svx0qL} yields the desired equation:
\begin{align}\label{eq:vSi5wvjUZy}
    G^2(z)&=-\frac{1}{N'}G'(z)+\frac{8}{\lambda }\left[G(z)-\frac{1}{z}\right].
\end{align}
In the large $N'$ limit, the first term goes to zero and the remaining quadratic equation in $G(z)$ has the simple solution:
\begin{align}\label{eq:SD5lzJwTlK}
    G(z)&=\frac{4}{\lambda}- \frac{4}{\lambda}\sqrt{1-\frac{\lambda}{2z}}\,,
\end{align}
where we have chosen the root that satisfies $G(z)\rightarrow 0$ for $z\rightarrow \infty$.  
Eqs.~(\ref{eq:SD5lzJwTlK}) and (\ref{eq:2GKGn244FH}) together then yield:
\begin{align}
    \sigma(u)=\frac{4}{\pi \lambda}\sqrt{\frac{\lambda}{2u}-1},
\end{align}
with $a=\frac{\lambda}{2}$. This result can be put into a more familiar form if we consider the expectation value of a function of the squares of the eigenvalues, $f(u)$. It is given by:
\begin{align}
    \braket{f}&=\int_0^{\frac{\lambda}{2}} f(u)\sigma(u) du=\int_0^{\sqrt{\frac{\lambda}{2}}}f(x^2)\sigma(x^2)2xdx=\int_{-\sqrt{\frac{\lambda}{2}}}^{\sqrt{\frac{\lambda}{2}}} f(x^2)\sigma(x^2)|x|dx.
\end{align}
This lets us identify
\begin{align}\label{eq:ZuxUCdvhWy}
    \rho(x)=|x|\sigma(x^2)=\frac{4}{\pi \lambda}\sqrt{\frac{\lambda}{2}-x^2}
\end{align}
as the distribution characterizing (the even moments of) the eigenvalues at the minimum of $U$. This is the well-known Wigner semicircle law with radius $\sqrt{\lambda/2}$.

We have focused on the saddle point of $SO(N)$ in the even (i.e., $N=2N'$) case, but the odd (i.e., $N=2N'+1$) case yields a similar analysis and the same eigenvalue density distribution at large $N'$. In particular, the eigenvalue potential energy in Eq.~\eqref{eq:hoe7tK35y6} picks up the additional term $-\frac{1}{N'^2}\sum_{i=1}^{N'}\log(\gamma_i^2)$ due to the factor $\prod_i \gamma_i^2$ in the measure in Eq.~\eqref{eq:tj14e0SXd7}. Consequently, no eigenvalue $\gamma_i$ can sit at the origin in the large $N'$ limit, and Eq.~\eqref{eq:cHnPl8iYIA} is replaced by\footnote{\label{ftnt:1} Eqs.~\eqref{eq:cHnPl8iYIA} and \eqref{eq:PQpXfWL7AJ} are more similar than they first appear. One of the eigenvalues in Eq.~\eqref{eq:cHnPl8iYIA} (say, $\gamma_{N'}$) is zero, since otherwise we may divide both sides by $\gamma_i$ and sum over $i$ to yield $4N'/\lambda=0$, a contradiction. Separating the zero eigenvalue from the others, Eq.~\eqref{eq:cHnPl8iYIA} reduces to
\begin{align}
    \frac{4}{\lambda}=\frac{1}{N'}\sum_{j=1,j\neq i}^{N'-1}\frac{1}{\gamma_i^2-\gamma_j^2}+\frac{1}{N'\gamma_i^2},\hspace{1cm}\text{ for }i=1,\ldots,N'-1.
\end{align}
Up to a factor of $1/2$ in the second term on the right hand side, this is the same as Eq.~\eqref{eq:PQpXfWL7AJ}.} 
\begin{align}\label{eq:PQpXfWL7AJ}
    \frac{4}{\lambda}=\frac{1}{N'}\sum_{j\neq i}\frac{1}{\gamma_i^2-\gamma_j^2}+\frac{1}{2N'\gamma_i^2},\hspace{1cm}\text{ for }i=1,\ldots,N'.
\end{align}
Although the details are slightly different, we can use Eq.~\eqref{eq:PQpXfWL7AJ} to derive a quadratic equation for the resolvent for the odd case just like in the even case, and the result is:
\begin{align}\label{eq:f07Yt2QTiz}
    G^2(z)=-\frac{1}{N'}\left[G'(z)+\frac{G(z)}{z}\right]+\frac{8}{\lambda}\left[G(z)-\frac{1}{z}\right].
\end{align}
Unsurprisingly, $G(z)$ and $\sigma(u)$ have the same large $N'$ behavior for the even and odd cases.%\footnote{It might be interesting to examine $1/N'$ corrections to the behavior of the saddle point for even and odd $N$ using Eqs.~\ref{eq:vSi5wvjUZy} and \ref{eq:f07Yt2QTiz}.  }

Having arrived at the eigenvalue density distribution, Eq.~\eqref{eq:ZuxUCdvhWy}, we can simplify the expectation value in Eq.~\eqref{eq:P2iNDsfwDy} for any function of the form $f=g\left(\frac{1}{N'}\sum_{i=1}^{N'}h(\gamma_i^2)\right)$. As long as the large $N'$ behavior of $f$ is subleading compared to $e^{-N'^2 U}$, the leading term in the asymptotic $1/N'$ expansion is:
%\footnote{If the function $g(x)$ is independent of $N'$, then Eq.~\eqref{eq:JIR6ANIZrF} just says that the $N'\to \infty$ limit of the left hand side is equal to the right hand side.}
\begin{align}\label{eq:JIR6ANIZrF}
    \left\langle g\left(\frac{1}{N'}\sum_{i=1}^{N'}h(\gamma_i^2)\right)\right\rangle&\sim g\left(\int \rho(x) h(x^2)dx\right),\hspace{1cm}N'\to \infty.
\end{align}
We must emphasize that the eigenvalue distribution in Eq.~\eqref{eq:ZuxUCdvhWy} does not correctly characterize the odd moments of the eigenvalues at the saddle points. Therefore, Eq.~\eqref{eq:JIR6ANIZrF} cannot be generalized by replacing $h(\gamma_i^2)$ and $h(x^2)$ by $h(\gamma_i)$ and $h(x)$ for any even or odd function $h(x)$.
\begin{comment}
\footnote{It is easy to find an example where Eq.~\eqref{eq:JIR6ANIZrF} fails to hold if $h(x^2)$ is replaced by an odd function $h(x)$. Let $g(x)=x^2$ and $h(x)=x$. Then the right hand side of Eq.~\eqref{eq:JIR6ANIZrF} evaluates to zero. By contrast, we can expand the left hand side of Eq.~\eqref{eq:JIR6ANIZrF} as
\begin{align}\label{eq:cczf3A6H5G}
    \frac{1}{(N')^2}\left\langle \sum_{i=1}^{N'}\gamma_i^2+\sum_{1\leq i<j\leq N'}\gamma_i\gamma_j\right\rangle= \frac{1}{(N')^2}\sum_{i=1}^{N'}\langle\gamma_i^2\rangle\sim \frac{1}{N'}\int x^2\rho(x)dx=\frac{\lambda}{8}\frac{1}{N'}
\end{align}
where the second equality follows because $\braket{\gamma_i\gamma_j}=0$ for $i\neq j$ due to the $\gamma_i\to -\gamma_i$ invariance of Eq.~\eqref{eq:P2iNDsfwDy}. Evidently, Eq.~\eqref{eq:JIR6ANIZrF} fails to capture the leading behavior of Eq.~\eqref{eq:cczf3A6H5G} at large $N'$. A more egregious example is the application of the semicircle law to the spinor Wilson loop; see Footnote~\ref{footnote:7a2ga2424gsdf4}.}
\end{comment}

Derivations of the semicircle distribution of the eigenvalues of the random antisymmetric Hermitian matrix can be found for instance in \cite{ashok2003unoriented,mehta1968distribution,mehta1983some}. The derivation in \cite{ashok2003unoriented} also implements the saddle point method and derives a ``loop equation'' \`a la Eq.~\eqref{eq:vSi5wvjUZy} for a resolvent closely related to our $G(z)$ in Eq.~\eqref{eq:u56ww2j0Wa}. On the other hand, the derivation in \cite{mehta1968distribution} proceeds by evaluating $N'-1$ of the integrals in Eq.~\eqref{eq:P2iNDsfwDy} (excluding, say, the $\gamma_1$ integral) using the method of orthogonal polynomials. The resulting expression for the distribution of $\gamma_1$ involves a sum of Hermite polynomials whose leading term in the asymptotic expansion in $1/N'$ yields the semicircle distribution. 
%There is no discussion of the difference in behavior of the even and odd moments of the eigenvalues in \cite{ashok2003unoriented,mehta1968distribution,mehta1983some}, an omission that is justified by the fact that any reasonable function of the eigenvalues of an antisymmetric Hermitian matrix would be invariant under $\gamma_i\to -\gamma_i$ (and thus an even function of the eigenvalues) because the eigenvalues of the $\Gamma$ matrices in Eqs.~\ref{eq:VcM3skO8Vn} and \ref{eq:tj14e0SXd7} come in pairs, $\pm \gamma_i$. We have chosen to discuss the even/odd discrepancy directly to provide a little more transparency.

As a check of the semicircle law (in particular of factors of 2), we note that Eq.~\eqref{eq:JIR6ANIZrF} implies that the expectation value of the $SO(N)$ Wilson loop in the fundamental representation is (choosing even $N$ for simplicity, but the large $N$ limit in the odd case is identical):
\begin{align}
\frac{1}{N}\braket{W_F}&=\biggr\langle\frac{2}{N}\sum_{i=1}^{N/2}\cosh{\gamma_i}\biggr\rangle=\frac{4}{\pi \lambda}\int_{-\sqrt{\lambda/2}}^{\sqrt{\lambda/2}} dx\sqrt{\frac{\lambda}{2}-x^2}\cosh{x}=\frac{2\sqrt{2}}{\sqrt{\lambda}}I_1\left(\sqrt{\frac{\lambda}{2}}\right).
\end{align}
This is in precise agreement with Eq.~\eqref{eq:1MQjpdsJek} at leading order at large $N$.

\subsection{Spinor representation}\label{sec:E4dda6KwCr}
The representation that primarily interests us in the present paper is the spinor representation of $SO(N)$, which we denote $R=sp$. Using the Weyl character formulas for $SO(2N')$ and $SO(2N'+1)$ (see e.g. \cite{balantekin2002character}), one finds 
\begin{align}\label{eq:XYlnrCkWg1}
\text{Tr}_{sp^{\pm}}(e^\Gamma)=2^{N'-1}\left(\prod_{i=1}^{N'}\cosh\left(\frac{\gamma_i}{2}\right)\pm \prod_{i=1}^{N'}\sinh\left(\frac{\gamma_i}{2}\right)\right)
\end{align}
when $N=2N'$ and
\begin{align}\label{eq:NMt3ZPcySa}
\text{Tr}_{sp}(e^\Gamma)=2^{N'}\prod_{i=1}^{N'}\cosh\left(\frac{\gamma_i}{2}\right)
\end{align}
when $N=2N'+1$. The two choices of sign for even $N$ correspond to the chiral and antichiral (Weyl spinor) representations of $SO(2N)$. One may also consider a non-chiral (Dirac spinor) representation by taking the sum of the two Weyl representations with opposite chirality. 

The method of orthogonal polynomials reduces the Wilson loop expectation value in the spinor representation to a determinant of an $N'\times N'$ matrix of associated Laguerre polynomials: $\braket{W_{sp}}=\text{dim}(sp)\times |D_{ij}|$, where
\begin{align}
    \text{dim }(sp)&=2^{N'-1},&D_{ij}&=L_{2i-2}^{2j-2i}\left(-\frac{\lambda}{32N}\right)e^{\frac{\lambda}{64N}},&&\text{for $N=2N'$ and}\label{eq:QqCzXFNc5L}
    \\
    \text{dim }(sp)&=2^{N'}, &D_{ij}&=L_{2i-1}^{2j-2i}\left(-\frac{\lambda}{32N}\right)e^{\frac{\lambda}{64N}},&&\text{for $N=2N'+1$.}\label{eq:2IynToBBl6}
\end{align}
We cite these results, derived in \cite{fiol2014exact}, to illustrate that extracting the large $N$ limit of the Wilson loop expectation value in the spinor representation directly using the orthogonal polynomial method is more involved than in the case of the fundamental representation. Therefore, we turn to the saddle point method and apply Eq.~\eqref{eq:JIR6ANIZrF} to Eq.~\eqref{eq:25ItjrOuK2}, given the characters in Eqs.~(\ref{eq:XYlnrCkWg1}) and (\ref{eq:NMt3ZPcySa}).

Without loss of generality, let us examine the odd case, $N=2N'+1$. We find:
\begin{align}
    \frac{1}{2^{N'}}\braket{W_{sp}}&=\frac{1}{2^{N'}Z'}\int \left(\prod_{i=1}^{N'}d\gamma_i\right) 2^{N'}\prod_{j=1}^{N'}\left(\cosh{\frac{\gamma_j}{2}}\right)e^{-N'^2U(\gamma_1,\ldots,\gamma_N')}\label{eq:kxOGvdG6gS}\\&\sim \text{exp}\left(N'\int_{-\sqrt{\lambda/2}}^{\sqrt{\lambda/2}}\rho(x)\log\left(\cosh\left(\frac{x}{2}\right)\right)dx\right).\label{eq:2XlJLe6dX6}
\end{align}
To get to the second line, we rewrote $\prod_{i=1}^{N'}\cosh(\frac{\gamma_i}{2})=\text{exp}\left(N'\times \frac{1}{N'}\sum_{i=1}^{N'}\log\left(\cosh\left(\frac{\gamma_i}{2}\right)\right)\right)$ and applied Eq.~\eqref{eq:JIR6ANIZrF}.
%\footnote{In \S~\ref{sec:J6TEzJQSXU} we stressed the difference in behavior of the even and odd moments of the eigenvalues at the saddle point in order to preempt the following incorrect application of Eq.~\eqref{eq:JIR6ANIZrF} to Eq.~\eqref{eq:kxOGvdG6gS}. ``First, the $\gamma_i\to -\gamma_i$ symmetry of Eq.~\eqref{eq:kxOGvdG6gS} lets us replace $\prod_{i=1}^{N'}\cosh\left(\frac{\gamma_i}{2}\right)$ by $\prod_{i=1}^{N'}e^{\frac{\gamma_i}{2}}=\text{exp}\left(\frac{1}{2}\sum_{i=1}^{N'}\gamma_i\right)$. Using the saddle point distribution, Eq.~\eqref{eq:ZuxUCdvhWy}, to replace $\frac{1}{N'}\sum_{i=1}^{N'}\gamma_i$ by $\int x\rho(x)dx=0$, we conclude that $\frac{1}{2^{N'}}\braket{W}=1$ in the large $N'$ limit.'' The second step is incorrect because it uses Eq.~\eqref{eq:JIR6ANIZrF} to determine the leading large $N'$ behavior of the expectation value of an \textit{odd} function of the  eigenvalues.\label{footnote:7a2ga2424gsdf4}} 
Substituting the explicit form for $\rho(x)$ from Eq.~\eqref{eq:ZuxUCdvhWy}, we 
find
%\begin{align}
%\int_{-\sqrt{\lambda/2}}^{\sqrt{\lambda/2}}\rho(x)\log\left(\cosh\left(\frac{x}{2}\right)\right)dx
%&=\frac{4}{\pi}\int_0^1 %\sqrt{1-u^2}\log\left(\cosh\left(\sqrt{\frac{\lambda}{8}}u\right)\right)du.\label{eq:PWl70hirur}
%\end{align}
that the expectation value of the spinor Wilson loop in the large $N$ limit is given by
\begin{equation}
 \frac{1}{2^{N'}}\braket{W_{sp}} \sim 
\exp\left[N' \frac{4}{\pi}\int_0^1 \sqrt{1-u^2}\log\left(\cosh\left(\sqrt{\frac{\lambda}{8}}u\right)\right)du\right]
\label{eq:PWl70hirur}
\end{equation}
This expression is valid at finite $\lambda$. We may expand it at small $\lambda$ using $\log(\cosh(x))= x^2/2-x^4/12+\ldots$ for small $x$, from which we get
\begin{align}
\frac{1}{N}\log\left[\frac{1}{2^{N'}}\braket{W}_{sp}\right]
&=\left(\frac{\lambda}{128}-\frac{\lambda^2}{12288}+O(\lambda^3)\right)+O(1/N).\label{eq:ef43efd3334432}
\end{align}
We can compare this with the leading order result from perturbation theory, Eq.~(\ref{MMpert}). For $G=SO(N)$, the spinor quadratic Casimir is $C_2(sp)=C(F)\frac{N(N-1)}{16}$ (see the Appendix). Keeping only the leading order terms in $g_{YM}$ and $1/N$, we thus find from (\ref{MMpert}) $\braket{W}_{sp}/\text{dim}(sp)=1+N\lambda/128(1+O(1/N))+O(g_{YM}^4)$. This matches the expansion in Eq.~\eqref{eq:ef43efd3334432} in the regime where $\lambda\ll 1$, $N\gg 1$ and $N\lambda\ll 1$ (despite the different order of limits potentially yielding, a priori, different results).

Extracting the leading terms of the expansion of Eq.~\eqref{eq:PWl70hirur} at strong coupling involves a little more work. First, 
we write $\log\left(\cosh\left(\sqrt{\frac{\lambda}{8}}u\right)\right)=\sqrt{\frac{\lambda}{8}}u-\log{2}+\log\left(1+e^{-\sqrt{\frac{\lambda}{2}}u}\right)$. The integrals over the first two terms can be evaluated explicitly. To expand the integral of the third term in $1/\sqrt{\lambda}$, we use $\log(1+e^{-\sqrt{\frac{\lambda}{2}}u})=\sum_{n=1}^\infty \frac{(-1)^{n+1}}{n} e^{-n\sqrt{\frac{\lambda}{2}}u}$, $\sqrt{1-u^2}=1-u^2/2-u^4/8+\ldots$, and $\int_0^1 u^m e^{-au}du=\Gamma(m+1)a^{-m-1}+O(e^{-a})$. 
%We also need to evaluate alternating sums of the form $\sum_{n=1}^\infty (-1)^{n+1}/n^{2k}$; for the lowest order correction corresponding to $k=1$, this evaluates to $\pi^2/12$. 
Combining these analytic results, we find:
\begin{align}
\log\left[\frac{1}{2^{N'}}\braket{W}_{sp}\right]
&=N\left(\frac{1}{3\pi}\sqrt{\frac{\lambda}{2}}-\frac{1}{2}\log{2}
+\frac{\pi}{6}\sqrt{\frac{2}{\lambda}}+O(\lambda^{-3/2},e^{-\sqrt{\lambda/2}})\right)+O(N^0).
\end{align}
We are especially interested in the leading term in the large $N$, large $\lambda$ expansion of the spinor $SO(N)$ Wilson loop, which is then given by 
\begin{align}\label{eq:lmZcUEVbH9}
    \frac{1}{\text{dim}(sp)}\braket{W_{sp}}\sim \exp\left(\frac{N}{3\pi}\sqrt{\frac{\lambda}{2}}\right).
\end{align}
We will compare Eq.~\eqref{eq:lmZcUEVbH9} to the dual D-brane calculation in \S~\ref{sec:HReGcbyOsR}. Note that, while we specialized to even $N$ for simplicity in the calculation above, the behavior (\ref{eq:lmZcUEVbH9}) applies to both even and odd $N$. 

\subsubsection*{Defect fermion description}
Before moving on to discuss the Wilson loop in the large rank (anti)symmetric representations, we point out a description of the spinor Wilson loop in terms of auxiliary ``defect" fermions living on the circle. This is similar to 
the known description of the $SU(N)$ Wilson loops in the symmetric and antisymmetric representations \cite{gomis2006holographic,hoyos2018defect}. 
%it is possible to describe the spinor Wilson loop by a one-dimensional effective defect action. 
In our case of spinor representation, the defect theory consists of real (Majorana) fermions $\chi^i$, $i=1,\ldots,N$, transforming in the fundamental representation of $SO(N)$ and coupled to $iA+\Phi^6$ (we choose $\Phi^6$ to be the scalar that couples to the half-BPS Wilson loop). The worldline action is given by:
\begin{align}\label{eq:A2i3r3gCB5}
    S_{\text{defect}}=\frac{i}{2}\int_0^1 dt \left[\chi^i \dot{\chi}^i+\chi^i \Gamma_{ij} \chi^j\right],
\end{align}
where we write $\Gamma=iA+\Phi^6$ for short-hand and, by suitable gauge-fixing, we may take $\Gamma$ to be constant on the Wilson loop. Consider the even $N$ case first. Upon canonical quantization, the fermion operators satisfy the Clifford algebra $\{\chi^i,\chi^j\}=\delta^{ij}$ and their Hilbert space is the $2^{\frac{N}{2}}$-dimensional space of the spinor representation. We can explicitly check that the contribution of the defect action, Eq.~\eqref{eq:A2i3r3gCB5}, to the partition function yields the characters of the spinor representation. Using an orthogonal transformation on the $\chi^i$ fields, we may write $\Gamma_{ij}=\text{diag}\left(\gamma_1\sigma_y,\gamma_2\sigma_y,\ldots,\gamma_{N/2}\sigma_y\right)$, and therefore:
\begin{align}
    Z_{\text{defect}}&=\int \prod_{i=1}^N D\chi_i e^{-S_{\text{defect}}}\nonumber\\&=\prod_{i=1}^{N/2}\int D\chi_{2i-1} D\chi_{2i}\text{exp}\left(\frac{1}{2}\int_0^1 dt (\begin{array}{cc}\chi^{2i-1} & \chi^{2i}\end{array})\left(\begin{array}{cc}i\frac{d}{dt} & \gamma_i \\ -\gamma_i & i\frac{d}{dt}\end{array}\right)\left(\begin{array}{c}\chi_{2i-1}\\ \chi_{2i}\end{array}\right)\right)\nonumber\\&=\prod_{i=1}^{N/2}\text{Det}^{\frac{1}{2}}\left(\begin{array}{cc}i\frac{d}{dt} & \gamma_i \\ -\gamma_i & i\frac{d}{dt}\end{array}\right)=\prod_{i=1}^{N/2}\prod_{n=0}^\infty (2n+1)^2\pi^2\left(1+\frac{\gamma_i^2}{(2n+1)^2\pi^2)}\right).\label{eq:XeQkedCl2d}
\end{align}
The product over odd integers, $2n+1$, appears because the functional determinant is taken over antiperiodic functions on the unit interval, $f_n(t)\propto e^{i(2n+1)\pi t}$. The functional determinant can be zeta-function regularized to yield:
\begin{align}
Z_{\text{defect}}&=2^{\frac{N}{2}}\prod_{i=1}^{\frac{N}{2}}\cosh\left(\frac{\gamma_i}{2}\right).
\end{align}
This is precisely the sum of the characters of the two chiral representations of $SO(N)$ for even $N$, given in Eq.~\eqref{eq:XYlnrCkWg1}. If one wanted to extract the chiral characters separately, one could compute the functional determinant over suitable linear combinations of periodic and antiperiodic boundary conditions on the circle.\footnote{This is because the path integral with periodic boundary conditions computes the trace with an extra insertion of $(-1)^{\cal F}$, where ${\cal F}$ is the fermion number. In terms of the 
Dirac matrices representation, $(-1)^{\cal F}$ is equivalent to the chirality matrix, and hence $(1\pm (-1)^{\cal F})/2$ projects onto the chiral and antichiral Weyl spinor representations.} 

The case of odd $N$ yields Eq.~\eqref{eq:XeQkedCl2d} with one additional path integral over $\chi^N$ with action $S=\frac{i}{2}\int dt \chi^N\dot{\chi}^N$. By suitable normalization of the path integral, the additional factor should be set to unity because the one-dimensional Clifford algebra $\{\chi,\chi\}=1$ defines a trivial spinorial representation of $SO(N)$. It follows that the defect theory also reproduces the character of $SO(N)$ for odd $N$, given in Eq.~\eqref{eq:NMt3ZPcySa}.

The defect theory defining the spinor Wilson loop differs from the defect theory defining the (anti)symmetric Wilson loops \cite{gomis2006holographic} in that the fermions are real and there is no additional gauge field on the worldline, which is necessary in the (anti)symmetric to fix the rank of the representation. The absence of the worldline gauge field should be reflected on the boundary conditions for the worldvolume gauge field on the dual D5-brane (the gauge field vanishes on the classical solution discussed in \S~\ref{sec:HReGcbyOsR} below).
%the fact that the wrapped D5-brane dual to the spinor Wilson loop--- which we shall discuss in \S~\ref{sec:HReGcbyOsR}--- does not carry fundamental string charge. 

\subsection{Large rank (anti)symmetric representation}\label{sec:cAhVgwXzVQ}
We next examine the Wilson loop in the antisymmetric and symmetric representations with large rank, $k\sim N$, which we denote $A_k$ and $S_k$, respectively. The generating functions for the characters of the antisymmetric and symmetric representations of $SO(N)$ take simple forms and can be used to determine the corresponding Wilson loops using the saddle point approximation in the usual 't Hooft limit with, additionally, $f=\frac{k}{N}$ held fixed. The analysis is nearly identical to the one for $SU(N)$ carried out in \cite{hartnoll2006higher}.

We recall from \cite{hartnoll2006higher} that, if $H$ is a Hermitian matrix with eigenvalues $h_i$, then the trace of $e^H$ in the rank $k$ antisymmetric representation of $SU(N)$ is given by $\text{tr}_{A_k}[e^H]=\frac{1}{2\pi i}\oint dt \frac{F_A(t)}{t^{N-k+1}}$,
where the antisymmetric generating function is $F_A(t)=\text{det}(1+te^H)=\prod_{i=1}^N (1+te^{h_i})$. Since antisymmetric Hermitian matrices are a subset of Hermitian matrices for which the eigenvalues come in pairs (e.g., $h_i=-h_{i+N/2}$ for $i=1,\ldots,N/2$ for the even $N$ case and $h_i=-h_{i+(N-1)/2}$ for $i=1,\ldots,(N-1)/2$ and $h_{N}=0$ for the odd $N$ case), the formula for the characters of $SO(N)$ in the rank $k$ antisymmetric representation is essentially the same:
\begin{align}\label{eq:ayocm27rSc}
    \text{Tr}_{A_k}[e^K]=\frac{1}{2\pi i}\oint dt \frac{F_A(t)}{t^{N-k+1}}=\frac{1}{2\pi i}\oint dt \frac{F_A(t)}{t^{k+1}},
\end{align}
where
\begin{align}\label{eq:GluClVNP3n}
    F_A(t)=\text{det}(1+te^K)=\prod_{i=1}^{N'} (1+te^{k_i})(1+te^{-k_i})\left\{\begin{array}{cc}1+t & \text{$N$ is odd}\\ 1 & \text{$N$ is even}\end{array}\right..
\end{align}
Note that $F_A(t)$ is a symmetric polynomial (i.e., $F_A(t)=t^NF_A(t^{-1})$), which justifies the second equality in Eq.~\eqref{eq:ayocm27rSc}. 

Similarly, the character of $SU(N)$ in the rank $k$ symmetric representation is $\text{tr}_{S_k}[e^H]=\frac{1}{2\pi i}\oint dt \frac{F_S(t)}{t^{k+1}}$, where the symmetric generating function is $F_S(t)=\text{det}(1-te^H)^{-1}=\prod_{i=1}^N (1-te^{h_i})^{-1}$ \cite{hartnoll2006higher}. We can again straightforwardly obtain the formula for the characters of $SO(N)$ in the rank $k$ symmetric representation from the character formula for $SU(N)$, with one modification (i.e., an extra factor of $1-t^2$ in the contour integral) related to the fact that the irreducible symmetric representations of $SO(N)$ must be traceless. Hence, the formula for the characters is:
\begin{align}\label{eq:oyaYloeXPY}
    \text{Tr}_{S_k}[e^H]&=\frac{1}{2\pi i}\oint dt \frac{(1-t^2)F_S(t)}{t^{k+1}},
\end{align}
where
\begin{align}\label{eq:Q7mtJgMdM9}
    F_S(t)&=\frac{1}{\text{det}(1-te^K)}=\prod_{i=1}^{N'} \frac{1}{(1-te^{k_i})(1-te^{-k_i})}\left\{\begin{array}{cc}(1-t)^{-1} & \text{$N$ is odd} \\ 1 & \text{$N$ is even}\end{array}\right..
\end{align}
One can check Eqs.~(\ref{eq:ayocm27rSc})-(\ref{eq:Q7mtJgMdM9}) by comparing with the character formulas for $SO(N)$ \cite{balantekin2002character}. 
%and we have verified that they indeed match for fixed choices of $k$ and $N$.

Let us use the generating functions and the saddle point analysis to determine the Wilson loop expectation value at large $N$ in the antisymmetric and symmetric representations. For simplicity, let us work in the even $N$ case. We first note that the two generating functions can be written compactly as:
\begin{align}
    F_{A,S}(t)&=\text{exp}\left(\pm \sum_{i=1}^{\frac{N}{2}}\left(\log(1\pm te^{k_i})+\log\left(1\pm te^{-k_i}\right)\right)\right),
\end{align}
where the upper sign corresponds to $A$ and the lower sign to $S$. Since both are dominated by $e^{-N^2 U}$ at large $N$, $U$ being the eigenvalue potential in Eq.~\eqref{eq:hoe7tK35y6}, the expectation values of $F_A$ and $F_S$ at large $N$ are both determined by the Wigner semicircle law. 
%Furthermore, the term $-t^2$ in Eq.~\eqref{eq:oyaYloeXPY} is invisible at leading order in $\frac{1}{N}$. 
Thus, we find that the large $N$ expectation value of the Wilson loop in the antisymmetric or symmetric representations of rank $k=fN$ with $f$ fixed as $N\to \infty$ is given by:
\begin{align}\label{eq:}
    \braket{W}_{A_k,S_k}&=\frac{1}{2\pi i}\oint dt \frac{(1-t^2)^{\frac{1\mp 1}{2}}}{t^{k+1}}\text{exp}\left(\pm\frac{4N}{\pi \lambda}\int_{-\sqrt{\frac{\lambda}{2}}}^{\sqrt{\frac{\lambda}{2}}}dx \sqrt{\frac{\lambda}{2}-x^2}\log (1\pm te^x)\right).
\end{align}
By expanding this result at small $\lambda$ and evaluating the resulting integrals,\footnote{The contour integrals may be evaluated using $\oint \frac{dt}{2\pi i} \frac{(1+t)^b}{t^{a+1}}=\begin{pmatrix}b\\a\end{pmatrix}$ and 
$\oint \frac{dt}{2\pi i} \frac{1}{(1-t)^b t^{a+1}}=\begin{pmatrix}b+a-1\\a\end{pmatrix}$. Note also that $\text{dim}(A_k)=\begin{pmatrix}N\\k\end{pmatrix}$ and 
$\text{dim}(S_k) = \begin{pmatrix}N+k-1\\k\end{pmatrix}-\begin{pmatrix}N+k-3\\k-2\end{pmatrix}$.}
one may check that this is in agreement with (\ref{MMpert}), using the values of the quadratic Casimir collected in Appendix. 

Note that the contour integrals in (\ref{eq:}) are the same as the analogous ones appearing in \cite{hartnoll2006higher} with the replacement $\lambda\to \lambda/2$.\footnote{A small difference between the result in (\ref{eq:}) and those in \cite{hartnoll2006higher} is the measure factor $1-t^2$ which implements tracelessness of the symmetric representation. However, the contribution of the $t^2$ subtraction term is subleading at leading order at large $N$.} Consequently, to obtain the strong coupling behavior, we can simply borrow their results. Then, for the antisymmetric representation in the $\lambda\gg 1$ regime, one finds:
\begin{align}\label{eq:UxmoHnC0eB}
    \braket{W_{A_k}}\sim \text{exp}\left[\frac{2N}{3\pi}\sqrt{\frac{\lambda}{2}}\sin^3{\theta_k}\right],
\end{align}
where $\theta_k$ satisfies:
\begin{align}\label{eq:rMz0QFxbWq}
    \pi f=\theta_k-\sin{\theta_k}\cos{\theta_k}.
\end{align}
For the symmetric representation, the saddle point that dominates in the large $\lambda$ limit with $\kappa=\frac{f}{4}\sqrt{\frac{\lambda}{2}}$ fixed yields the result:
\begin{align}\label{eq:ULMUF3S64u}
    \braket{W_{S_k}}\sim \text{exp}\left[2N(\kappa\sqrt{1+\kappa^2}+\sinh^{-1}{\kappa}\right].
\end{align}
We will compare these results below to D3 and D5-branes in $AdS_5\times \RP^5$.

\section{Dual string theory in $AdS_5\times \RP^5$}\label{sec:jKYRKLXLoH}

As explained in \cite{witten1998baryons}, the holographic dual to ${\cal N}=4$ SYM with $SO(N)$ or $USp(N)$ gauge group is type IIB string theory on the $AdS_5\times \RP^5$ orientifold. Here $ \RP^5=S^5/\mathbb{Z}_2$, 
where the $\mathbb{Z}_2$ quotient acts by identification of antipodal points (the orientifolding procedure implies that as the string goes around a non-contractible cycle, its orientation is reversed). This background arises as 
the near-horizon limit of a stack of D3-branes placed at an orientifold plane in flat space. The difference between the orthogonal and symplectic groups corresponds to a choice of ``discrete torsion" for the Neveu-Schwarz and Ramond-Ramond 2-forms $B_{NS}$ and $B_{RR}$. Essentially, this corresponds to a discrete choice for the holonomy on a $\mathbb{RP}^2 \subset \mathbb{RP}^5$, $\exp(i\int_{\mathbb{RP}^2} B_{NS})=e^{2\pi i \theta_{NS}}=\pm 1$ and 
$\exp(i\int_{\mathbb{RP}^2} B_{RR})=e^{2\pi i \theta_{RR}}=\pm 1$. The choices 
$(\theta_{NS},\theta_{RR})=(0,0)$ and $(\theta_{NS},\theta_{RR})=(0,1/2)$ correspond respectively to the $SO(2N')$ and $SO(2N'+1)$ gauge theory, while the two choices with non-zero $\theta_{NS}$ are dual to the gauge theory with symplectic group. The discrete torsion will not play an explicit role in the calculations below, but it is important to note that the D5-brane dual to the spinor Wilson loop, which wraps a $\mathbb{RP}^4\subset \RP^5$, is only possible when $\theta_{NS}=0$ \cite{witten1998baryons}. This is in line with the fact that there are no spinor representations for $USp(N)$ group. 

Before moving on to compare the Wilson loop results to strings and branes in $AdS_5\times \RP^5$, let us first briefly review the relation between gauge theory and string theory parameters. The relevant part of the type IIB supergravity Lagrangian that involves the metric and self-dual RR 5-form is (we work in Euclidean signature throughout)
\begin{equation}
S = -\frac{1}{2\kappa_{10}^2} \int d^{10}x \sqrt{g}\left({\cal R}-\frac{1}{4\cdot 5!} (F_5)^2+\ldots \right)\,,
\end{equation}
where $2\kappa_{10}^2 = (2\pi)^7 g_s^2 \alpha'^4$. In the more familiar case of the duality with the $SU(N)$ gauge theory, the $AdS_5\times S^5$ background is given by 
\begin{equation}
ds^2 = L^2 \left(ds^2_{AdS_5}+ds^2_{S^5}\right)
\end{equation}
where $ds^2_{AdS_5}$ and $ds^2_{S^5}$ are the metrics on unit radius $AdS_5$ and $S^5$. The 5-form field strength is given by $F_5 = \frac{4}{L}(\omega_5+\tilde{\omega}_5)$, where $\omega_5$ and $\tilde{\omega}_5$ are volume forms on $AdS_5$ and $S^5$ with unit radius. The radius is fixed by the condition that the flux of the 5-form is 
$\frac{1}{2 \kappa_{10}^2 T_3}\int_{S^5} F_5 = N$, corresponding to $N$ D3-branes (here $T_3 = (2\pi)^{-3} (\alpha')^{-2} g_s^{-1}$ is the D3-brane tension). Using ${\rm vol}(S^5) =\pi^3 L^5$, this gives the relation $L^4 = 4\pi g_s \alpha'^2 N$. The string coupling $g_s$ is related to the $SU(N)$ SYM coupling normalized as in (\ref{eq:yhbwoLehbV}) by $g_{YM}^2 = 4\pi g_s$.    

Similarly, the $AdS_5\times \RP^5$ background relevant to $SO(N)$ and $USp(N)$ gauge theory takes the form 
\begin{equation}
ds^2 = L^2 \left(ds^2_{AdS_5}+ds^2_{\RP^5}\right)
\end{equation}
where the metric on $\RP^5$ is locally the same as $S^5$, but due to the $\mathbb{Z}_2$ quotient ${\rm vol}(\RP^5)=\frac{1}{2}{\rm vol}(S^5)$. The 5-form field strength still takes the form $F_5=\frac{4}{L}(\omega_5+\tilde{\omega}_5)$, but now the flux quantization condition for 
$SO(2N')$, $SO(2N'+1)$ and $USp(2N')$ requires  
\begin{equation}
\frac{1}{2 \kappa_{10}^2 T_3}\int_{\RP^5} F_5 = N'\,.
\label{RP-flux}
\end{equation}
This gives $L^4 = 8\pi g_s {\alpha'}^2 N'$. This relation is valid to leading order at large $N$. Taking into account also the contribution to the RR charge due to the orientifold three-plane,\footnote{The charge of the orientifold three-plane for the $SO(2N')$, $SO(2N'+1)$ and $USp(2N')$ cases is respectively $-1/4$, $+1/4$, $+1/4$, see e.g. Table 1 in \cite{Bergman:2001rp}. Adding this contribution to the right-hand side of (\ref{RP-flux}), one can write the result for the radius in terms of $N$ (rather than $N'$) as in (\ref{RP-radius}).} 
the formula for the radius can be written for $SO(N)$ and $USp(N)$ as \cite{blau1999subleading}
\begin{equation}
L^4 = 8\pi g_s{\alpha'}^2\left(\frac{N}{2}\pm \frac{1}{4}\right)\,.
\label{RP-radius}
\end{equation} 
The mapping between string coupling and SYM coupling normalized as in (\ref{eq:yhbwoLehbV}) is given by
\begin{equation}
g_{YM}^2 = 8\pi g_s\,.
\label{gYM-map}
\end{equation} 
Note the factor of 2 difference compared to the analogous relation in the $SU(N)$ case.\footnote{For an explanation of this factor of 2 difference between $U(N)$ and $SO(N)$ gauge theory on D-branes, see for instance section 13.3 of \cite{Polchinski:1998rr}.} 

As a check of the dictionary reviewed above, let us briefly recall the matching of the conformal anomalies \cite{Henningson:1998gx, Balasubramanian:1999re, blau1999subleading}. A quick way to extract the $a$-anomaly coefficient from the holographic dual is to evaluate the on-shell action on the background solution, using the hyperbolic ball coordinates on $AdS_5$ where the boundary is a sphere $S^4$. The conformal anomaly then manifests itself in the fact that the volume of $AdS_{d+1}$ with even $d$ is logarithmically divergent (see e.g. \cite{Diaz:2007an, Casini:2011kv}) 
\begin{equation}
{\rm vol}(AdS_{d+1}) = \frac{2(-\pi)^{d/2}}{\Gamma\left(1+\frac{d}{2}\right)} \log R
\label{volAdS}
\end{equation}  
where $R$ may be identified with the radius of the boundary sphere. Integrating over the $\RP^5$ directions yields the action
\begin{equation}
S = -\frac{L^5}{2\kappa_{10}^2}{\rm vol}(\RP^5) \int_{AdS_5} d^5 x\sqrt{g}\left({\cal R}_5+\frac{12}{L^2}\right) 
\end{equation}
where ${\rm vol}(\RP^5)=\pi^3/2$ is the volume of $\RP^5$ with unit radius. 
Using (\ref{RP-radius}) and (\ref{volAdS}) this yields \cite{blau1999subleading}
\begin{equation}
S = \left(\frac{N(N\pm 1)}{2}+\frac{1}{8}\right)\log R\,.
\end{equation}
The holographic prediction for the $a$-anomaly coefficient (normalized so that $a=1/90$ for a free massless scalar) is then $a=\frac{N(N\pm 1)}{2}+O(N^0)$, in agreement, up to order $N$, with the anomaly of the dual $SO(N)$ and $USp(N)$ SYM theories.\footnote{The $a$-anomaly in ${\cal N}=4$ SYM with group $G$ is given by $a=\frac{1}{90}\left(6+4\cdot \frac{11}{2} + 62\right){\rm dim}(G) ={\rm dim}(G)$.} Note that the term of order $N^0$ is not expected to match from this classical calculation, because there can be corrections of order $N^0$ coming from supergravity one-loop effects. It would be interesting to compute them, along the lines of what was done in \cite{Beccaria:2014xda} for the $SU(N)$ case, and check whether they agree with the exact gauge theory answer, $a=\frac{N(N\pm 1)}{2}$. 

In the remainder of this section, we will focus exclusively on the $SO(N)$ case, and compare the localization prediction for the Wilson loop with the dual strings and D-branes in $AdS_5\times \RP^5$. We will work in the leading large $N$ limit, where the finite shift in (\ref{RP-radius}) will not play a role.
 
\begin{comment}
In the present section, we examine the $AdS_5\times \RP^5$ duals to the spinor, large rank antisymmetric and large rank symmetric Wilson loops. As prescribed in \cite{witten1998baryons}, the $SO(N)$ Wilson loop in the spinor representation is dual to a neutral D5-brane wrapped on an $\RP^4$ submanifold of $\RP^5$. We confirm this prescription explicitly by comparing the D5-brane's action to Eq.~\eqref{eq:lmZcUEVbH9}. Meanwhile, the duals to the $SO(N)$ Wilson loop in the large rank antisymmetric and symmetric representations are almost identical to the duals to the $SU(N)$ Wilson loop in the corresponding representations \cite{drukker2005all,yamaguchi2006wilson,gomis2006holographic,hartnoll2006higher}. Specifically, the symmetric Wilson loop is dual to the charged D3-brane in $AdS_5$ studied in \cite{drukker2005all}, except with the electric charge of the brane suitably rescaled, which we confirm by comparing the D3-brane action to Eq.~\eqref{eq:ULMUF3S64u}. Furthermore, the antisymmetric Wilson loop is dual to the charged D5-brane studied in \cite{yamaguchi2006wilson}, except it is now wrapped on $S_4$ submanifolds of $\RP^5$, which we confirm by comparing the D5-brane action to Eq.~\eqref{eq:UxmoHnC0eB}. 
\end{comment}

\subsection{Fundamental string}

As is well-known, the dual to the Wilson loop in the fundamental representation is a string worldsheet corresponding to a minimal surface ending on the Wilson loop contour on the boundary \cite{maldacena1998wilson}. The fundamental string action is 
\begin{equation}
S_{F1} = T_{F1} \int d^2\sigma \sqrt{h}\,,\qquad T_{F1}=\frac{1}{2\pi\alpha'} 
\end{equation}
where $h=\det h_{\alpha\beta}$ and $h_{\alpha\beta}$ is the induced metric. 

For a Wilson loop with constant scalar coupling, the string surface is pointlike in the internal $\RP^5$ space and extends entirely within $AdS_5$. In the case of the 1/2-BPS Wilson loop supported on a circle or infinite straight line, the minimal surface is well known \cite{drukker1999wilson}, and corresponds to an $AdS_2\subset AdS_5$. Specifically, using the Poincar\'e coordinates in $AdS_5$
\begin{align}\label{eq:Zmbgznmcs2}
    ds^2=\frac{L^2}{z^2}\left(dz^2+dx_\mu dx^\mu\right)
\end{align}
the minimal surface dual to a circular Wilson loop of radius $a$ in the $(x^1, x^2)$-plane may be written as
\begin{equation}
x^1 = \frac{a \cos\tau}{\cosh\sigma} \,,\qquad x^2 = \frac{a \sin\tau}{\cosh\sigma}\,,\qquad 
z = a \tanh\sigma
\end{equation}
where $0<\sigma<\infty$ and $0<\tau<2\pi$ are the worldsheet coordinates. 
This is a ``hemisphere" $z^2+(x^1)^2+(x^2)^2=a^2$ embedded in $AdS_5$. The induced worldsheet metric is 
\begin{equation}
ds^2_2 = \frac{L^2}{\sinh^2\sigma}\left(d\tau^2+d\sigma^2\right)
\end{equation}
which is the hyperbolic disk metric of $AdS_2$.\footnote{Similarly, for the 1/2-BPS Wilson loop on the infinite straight line, the string solution is $z=\sigma$, $x^1=\tau$, with $0<\sigma<\infty$, $-\infty<\tau<\infty$. This gives the Poincar\'e half-plane.} To compute the regularized action on this solution, one may either introduce a cutoff at $z=\epsilon$ and add a boundary term to remove the linear divergence proportional to the perimeter \cite{maldacena1998wilson, drukker1999wilson}, or directly use the well-known formula for the regularized volume of the hyperbolic disk (with unit radius)
\begin{equation}
{\rm vol}(AdS_2) = -2\pi\,. 
\end{equation}
Making use of this result, we get
\begin{equation}
S_{F1} = \frac{1}{2\pi\alpha'}\int d^2\sigma \sqrt{h} = \frac{L^2}{2\pi\alpha'}{\rm vol}(AdS_2)\,.
\end{equation} 
Using $L^4 = 8\pi g_s \alpha'^2 \frac{N}{2}$ for $SO(N)$ as reviewed above (we drop the finite shift in (\ref{RP-radius}) here and everywhere below), we find
\begin{equation}
S_{F1} = -\sqrt{4\pi g_s N} = -\sqrt{\frac{\lambda}{2}}\,, 
\label{F1-action}
\end{equation}
where in the last step we used the relation (\ref{gYM-map}) and $\lambda = g^2_{YM}N$. The strong coupling behavior of the fundamental Wilson loop is then
\begin{align}\label{eq:MjC3BGYJKQ}
   \frac{1}{N} \langle  W_F\rangle \sim e^{-S_{F1}} = e^{\sqrt{\frac{\lambda}{2}}}\,.
\end{align}
Comparing with the asymptotic expansion in $1/\lambda$ of the $O(N^0)$ term in Eq.~\eqref{eq:1MQjpdsJek}, $\frac{1}{N}\braket{W_F}\sim \sqrt{\frac{2}{\pi}}\left(\frac{2}{\lambda}\right)^{\frac{3}{4}}e^{\sqrt{\frac{\lambda}{2}}}$, we see that Eq.~\eqref{eq:MjC3BGYJKQ} reproduces the large $N$, large $\lambda$ behavior of the fundamental Wilson loop in $\mathcal{N}=4$ $SO(N)$ SYM. It would be interesting to study subleading corrections, especially the non-planar $1/N$ corrections given in (\ref{eq:1MQjpdsJek}). In particular, the term of order $1/N$, which is absent in the $SU(N)$ theory, should come from a worldsheet with a crosscap.

\subsection{The D5-brane dual to the spinor Wilson loop}\label{sec:HReGcbyOsR}

Let us now turn to the case of main interest in this paper-- namely, the Wilson loop in the spinor representation of $SO(N)$. As argued in \cite{witten1998baryons}, the dual object is a D5-brane wrapping a $\RP^4$ ``equator" in $\RP^5$ (see Figure.~\ref{subfig:1}), and having $AdS_2$ induced geometry within $AdS_5$, like the fundamental string of the previous section. As mentioned above, such wrapping is only possible when the NS discrete torsion vanishes, $\theta_{NS}=0$. Unlike the D-branes dual to the (anti)symmetric representations \cite{drukker2005all, Yamatsu:2015npn, gomis2006holographic}, on this D5-brane worldvolume there is no $U(1)$ gauge field turned on. The classical DBI action evaluated on the solution is then simply given by the volume of the D5-brane, which has $AdS_2\times \RP^4$ induced geometry. The on-shell action is then
\begin{equation}
S_{D5} = T_5 \int d^6\sigma \sqrt{h} = T_5 L^6 {\rm vol}(AdS_2){\rm vol}(\RP^4)\,.
\end{equation}
where ${\rm vol}(AdS_2)=-2\pi$ and ${\rm vol}(\RP^4)=\frac{4}{3}\pi^2$ are the volumes of $AdS_2$ and $\RP^4$ with unit radius. 
Here $T_5$ is the D5-brane tension, which reads\footnote{In general, for a Dp-brane, the tension is $T_p =(2\pi)^{-p} (\alpha')^{-(p+1)/2} g_s^{-1}$, see e.g. \cite{Johnson:2000ch}.}
\begin{equation}
T_5 = \frac{1}{(2\pi)^5 \alpha'^3 g_s} = \frac{N}{8\pi^4 L^6}\sqrt{\frac{\lambda}{2}}
\label{T5-tension}
\end{equation}
where we used (\ref{RP-radius}) and (\ref{gYM-map}). Hence we find
\begin{equation}
S_{D5} = -\frac{N}{3\pi}\sqrt{\frac{\lambda}{2}}\,,
\label{D5-spinor}
\end{equation}
or, in terms of the spinor Wilson loop expectation value
\begin{equation}
\frac{1}{2^{N/2}} \langle W_{sp} \rangle \sim e^{-S_{D5}} =\exp\left(\frac{N}{3\pi}\sqrt{\frac{\lambda}{2}}\right)\,.  
\end{equation}
This precisely matches the $\mathcal{N}=4$ SYM result predicted by localization, Eq.~\eqref{eq:lmZcUEVbH9}.  

\begin{comment}
To compute the supergravity regime of the spinor circular Wilson loop in the bulk, we need to work with a different classical stringy object whose boundary conditions are defined by the Wilson loop. The correct object, identified in \cite{witten1998baryons}, is the $D5-$brane with four dimensions wrapped on an $\RP^4$ subspace of $\RP^5$ (see Figure.~\ref{subfig:1}) and the remaining two dimensions forming the hemisphere in $AdS_5$ incident on the Wilson loop. Since the geometry of the D5-brane is $AdS_2\times \RP^4$, its regularized volume is $V_5=(-2\pi)\times \left(\frac{1}{2} \frac{8\pi^2}{3}\right)$. Furthermore, its tension $T_5$ is related to the fundamental string tension $T$ and coupling $g_s$ via \cite{Johnson:2000ch}
\begin{align}\label{eq:hqKDyDZolA}
    T_5&=\frac{T^3}{(2\pi)^2g_s}.
\end{align}
Then, the exponential of the regularized D5-brane action is 
\begin{align}\label{eq:clUWLPoMWa}
    e^{-S_5}=\text{exp}\left(\frac{1}{3(2\pi)^2g_s}\left(\frac{\lambda}{2}\right)^{\frac{3}{2}}\right),
\end{align}
which matches the $\mathcal{N}=4$ SYM, Eq.~\eqref{eq:lmZcUEVbH9}, result as long as $g_s=\frac{\lambda}{8\pi N}=\frac{g_{YM}^2}{8\pi}$. 
\end{comment}

By comparing Eqs.~(\ref{D5-spinor}) and (\ref{F1-action}), we see that the actions of the D5-brane dual to the circular Wilson loop in the spinor representation and the fundamental string dual to the circular Wilson loop in the fundamental representation are related by 
\begin{align}\label{eq:wBJNu5u8pA}
    S_{D5}=\frac{N}{3\pi} S_{F1}.
\end{align}
This relationship in fact holds for any Wilson loop that is $SO(5)$ invariant (i.e., for which $\Theta^I$ in Eq.~\eqref{eq:2W2phlg0Lb} is a point in $\RP^5$), regardless of the shape of the contour in $\mathbb{R}^4$. The reason is simple. If the minimal fundamental string surface in $AdS_5$ is $\Sigma$ and $A(\Sigma)$ is its area, then $S_{F1}=T_{F1} A(\Sigma)$. On the other hand, the volume of the minimal wrapped D5-brane dual to the spinor Wilson loop, which has induced geometry $\Sigma\times \RP^4$, is $A(\Sigma)\frac{4\pi^2 L^4}{3}$. Given $S_{D5}=T_5 A(\Sigma)\frac{4\pi^2 L^4}{3}$ and $T_5 L^6=\frac{N}{4\pi^3}T_{F1}L^2$ (see Eq.~\eqref{T5-tension}), then Eq.~\eqref{eq:wBJNu5u8pA} follows. This argument is similar to the one in \cite{hartnoll2006two}, which established an analogous universal relationship between the fundamental Wilson loop and the rank $k$ antisymmetric Wilson loop. The argument for the spinor Wilson loop is even simpler since the wrapped D5-brane dual to the spinor Wilson loop carries zero electric field and is governed simply by the Nambu-Goto action. One example of a useful consequence of Eq.~\eqref{eq:wBJNu5u8pA}, is that, since we know the quark-antiquark potential at strong coupling takes the form $V_F(R)=-2\pi\gamma T_{F1}L^2/R = -\gamma\sqrt{\frac{\lambda}{2}}\frac{1}{R}$ where $\gamma$ is a numerical factor that can be obtained from the relevant minimal surface \cite{maldacena1998wilson},\footnote{The explicit value is $\gamma = 4\pi^2\sqrt{2}/\Gamma\left(\frac{1}{4}\right)^4$.} and $R$ the distance between the quarks, then the potential between probe charges in the spinor representation must be $V_{sp}=-2\pi \gamma \frac{N}{3\pi}\frac{T_{F1}L^2}{R} =-\gamma \frac{N\sqrt{\frac{\lambda}{2}}}{3\pi}\frac{1}{R}$. 

\subsection{The D3/D5-branes dual to large rank (anti)symmetric Wilson loops}\label{sec:0pfY1VVBbg}

\begin{figure}[t!]
    \centering
    \begin{subfigure}[t]{0.5\textwidth}
        \centering
        \begin{tikzpicture}
            \draw (0,0) arc (180:0:2cm);
            \draw[dashed,blue] (0,0) arc (180:0:2cm and 0.75cm);
            \draw [blue] (0,0) arc (180:360:2cm and 0.75cm);
            \draw[<->] (0.5,-0.25) -- (3.5,0.25);
            \draw[<->] (1.5,0.6) -- (2.5,-0.6);
            \node at (2.3,2.3) {$\RP^5$};
            \node at (3.2,-0.3) {$\RP^4$};
        \end{tikzpicture}
        \caption{}
        \label{subfig:1}
    \end{subfigure}%
    ~ 
    \begin{subfigure}[t]{0.5\textwidth}
        \centering
        \begin{tikzpicture}
            \draw [dashed] (2,2) -- (2,0);
            \draw [dashed] (2,0) -- (3.414,1.414);
            \draw (0,0) arc (180:0:2cm);
            \draw[dashed] (0,0) arc (180:0:2cm and 0.75cm);
            \draw (0,0) arc (180:360:2cm and 0.75cm);
            \draw[blue,dashed] (0.586,1.414) arc (180:0:1.414cm and 0.3cm);
            \draw [blue] (0.586,1.414) arc (180:360:1.414cm and 0.3cm);
            \draw[<->] (0.5,-0.25) -- (3.5,0.25);
            \draw[<->] (1.5,0.6) -- (2.5,-0.6);
            \node at (2.3,2.3) {$\RP^5$};
            \node at (0.386,1.514) {$S^4$};
            \node at (2.2,0.5) {$\theta_k$};
        \end{tikzpicture}
        \caption{}
        \label{subfig:2}
    \end{subfigure}
    \caption{The real projective space $\RP^5$ can be visualized as the upper hemisphere of $S^5$ with antipodal points on the equator identified. Since the neutral D5-brane dual to the spinor Wilson loop wraps an $\RP^4$ submanifold of $\RP^5$, we can think of it as wrapping the equator, as is indicated by the blue curve in (a). Meanwhile, the D5-brane dual to the antisymmetric Wilson loop wraps an $S^4$ submanifold of $\RP^5$ at polar angle $\theta_k$, as is indicated by the blue curve in (b).}
\end{figure}
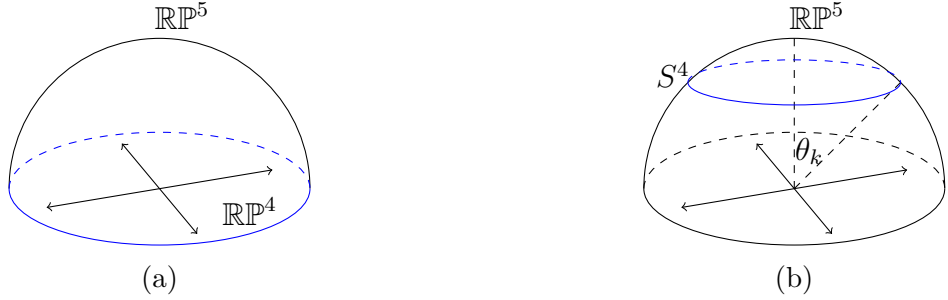

In \S~\ref{sec:cAhVgwXzVQ}, we found that the $SO(N)$ Wilson loops in the large rank symmetric and antisymmetric representations, in the supergravity regime, were equal to the corresponding $SU(N)$ Wilson loops up to the replacement $\lambda\to \lambda/2$. Consequently, we expect that the D3 and D5-branes dual to the $SO(N)$ Wilson loops are closely related to the corresponding branes dual to the $SU(N)$ Wilson loops, which were identified in \cite{hartnoll2006higher,drukker2005all,yamaguchi2006wilson,gomis2006holographic}. 

To review the basic facts about those D-brane solutions, it is convenient to use the following $AdS_2\times S^2$ slicing of the $AdS_5$ metric (see e.g.\cite{Giombi:2020amn} for more details on the form of the brane solutions and their actions using these coordinates)
\begin{equation}
ds^2_{AdS_5} = L^2 \left(du^2+\cosh^2u ds^2_{AdS_2}+\sinh^2u d\Omega_2^2 \right)\,.
\label{AdS2-slice}
\end{equation}
The D3-brane of \cite{drukker2005all} extends along the $AdS_2$ and $S^2$ directions within $AdS_5$, with $u=u_k$ constant, and it is point-like in the internal space $S^5$ (or, in our case, $\RP^5$). In addition, there is a non-zero worldvolume gauge field given by 
\begin{equation}
F = i\cosh u_k\, \omega_2
\end{equation}
where $\omega_2$ denotes the volume form on $AdS_2$. The parameter $u_k$ is fixed by requiring that the fundamental string charge is equal to $k$, the rank of the dual representation. This yields the condition, written in terms of D3-brane tension and radius of $AdS_5$ 
\begin{equation}
k= 2\pi \alpha' T_3 4\pi L^2 \sinh u_k = \frac{L^2 }{\pi \alpha' g_s} \sinh u_k
\label{kappa-eq}
\end{equation} 
We have written the result in this (perhaps unfamiliar-looking) way so that it can be easily applied to both the $AdS_5\times S^5$ and $AdS_5\times \RP^5$ cases, using the appropriate value of the $AdS_5$ radius. For the former case, using $L^4 = 4\pi g_s \alpha'^2 N = g_{YM}^2 N \alpha'^2$, (\ref{kappa-eq}) yields 
\begin{equation}
\sinh u_k = \frac{f\sqrt{\lambda}}{4}\,,\qquad f=\frac{k}{N}\,,
\end{equation}
while for the $AdS_5\times \RP^5$  dual to $SO(N)$ gauge group, using (\ref{RP-radius}) and (\ref{gYM-map}), we get 
\begin{equation}
\sinh u_k = \frac{f\sqrt{\lambda/2}}{4}\,,\qquad f=\frac{k}{N}\,.
\label{eq:mHU3lD1vo2}
\end{equation}
The D3-brane on-shell action, obtained by summing the DBI and Wess-Zumino contributions, and supplemented by a boundary term implementing a Legendre transform on the gauge field, yields \cite{drukker2005all} (see \cite{Giombi:2020amn} for the calculation in the coordinates used here)
\begin{equation}
S_{D3} = -4\pi^2 L^4 T_3 \left(u_k+\sinh u_k \cosh u_k\right) 
= -2N  \left(u_k+\sinh u_k \cosh u_k\right) 
\end{equation} 
In the $SO(N)$ case, using (\ref{eq:mHU3lD1vo2}), we see that this agrees with \eqref{eq:ULMUF3S64u}, i.e. it is the same as the $SU(N)$ case, up to the replacement $\frac{f\sqrt{\lambda}}{4N}\rightarrow  \frac{f\sqrt{\lambda/2}}{4N}$.

The $AdS_2\times S^4$ D5-brane \cite{Camino:2001at, yamaguchi2006wilson} dual to the antisymmetric Wilson loop occupies the $AdS_2$ in the metric (\ref{AdS2-slice}) with  $u=0$, and it wraps an $S^4\subset S^5$ at a constant polar angle $\theta=\theta_k$. Similarly to the D3-brane, there is a worldvolume gauge field with field strength components along $AdS_2$, given explicitly by
\begin{equation}
F = i \cos\theta_k \, \omega_2\,. 
\end{equation}
We can have the same solution in the $AdS_5\times \RP^5$ case. We may think of $\RP^5$ as the upper hemisphere of $S^5$ with antipodal points identified on the equator. The brane solution then wraps an $S^4$ ``latitude" at angle $\theta=\theta_k$ on the upper hemisphere, as shown in Figure~\ref{subfig:2}. The requirement that the flux of the worldvolume gauge field is equal to $k$ yields the condition 
\begin{equation}
k = 8\pi^3 \alpha' T_5 L^4 (\theta_k-\sin\theta_k \cos\theta_k)
= \frac{L^4}{4\pi^2 g_s \alpha'^2} (\theta_k-\sin\theta_k \cos\theta_k)\,.
\end{equation}
In terms of gauge theory parameters, this can be written, both for the $SU(N)$ and $SO(N)$ cases, as 
\begin{equation}
\theta_k-\sin\theta_k \cos\theta_k = \frac{\pi k}{N}\,.
\label{thetak}
\end{equation}
Note that in the $SO(N)$ case we can restrict $k \leq N'$, where $N'$ is the rank of $SO(N)$, which is $N'=N/2$ for $N$ even and $N'=(N-1)/2$ for $N$ odd. Then we see that it is consistent to restrict $\theta_k$ to the upper hemisphere.\footnote{The equatorial case $\theta_k=\frac{\pi}{2}$, corresponding to $k=\frac{N}{2}$ for even $N$, is a subtle special case because antipodal points on the equator are identified under the quotient $S^5/\mathbb{Z}_2\cong \RP^5$. The fact that the volume of $\RP^4$ is half that of $S^4$ may indicate that the D5-brane dual to the antisymmetric Wilson loop corresponding to $k=N/2$ is doubly wrapped, or alternatively that the singly-wrapped D5-brane is dual to one of the two irreducible, self-dual representations into which the $k=N/2$ antisymmetric representation may be decomposed.}
The total action of the D5-brane, including the relevant boundary terms, yields
\cite{yamaguchi2006wilson}
\begin{equation}
S_{D5} = -\frac{16\pi^3}{3} T_5 L^6 \sin^3\theta_k\,.
\end{equation}
For the $AdS_5\times \RP^5$ dual to the $SO(N)$ theory, we then have, using (\ref{T5-tension}):
\begin{equation}
S_{D5} = -\frac{2N}{3\pi}\sqrt{\frac{\lambda}{2}}\sin^3{\theta_k}\,,
\end{equation}
which matches the expectation value of the $SO(N)$ Wilson loop in the antisymmetric representation given in Eq.~\eqref{eq:UxmoHnC0eB}. Again, the difference with the $SU(N)$ theory result to this order is simply the replacement $\sqrt{\lambda}\rightarrow \sqrt{\lambda/2}$, in agreement with the localization analysis.  

\begin{comment}
$S_{\text{tot}}=-\frac{L^2}{\alpha'}\frac{2N}{3\pi}\sin^3{\theta_k}$. Translated into the parameters of $SU(N)$ $\mathcal{N}=4$ SYM, the D5-brane action becomes $S_{\text{tot}}=-\sqrt{\lambda}\frac{2N}{3\pi}\sin^3{\theta_k}$, which matches the expectation value of the $SU(N)$ Wilson loop in the antisymmetric representation \cite{yamaguchi2006wilson,hartnoll2006higher}. Translated into the parameters of $SO(N)$ $\mathcal{N}=4$ SYM, the D5-brane action becomes $S_{\text{tot}}=-\sqrt{\frac{\lambda}{2}}\frac{2N}{3\pi}\sin^3{\theta_k}$, which matches the expectation value of the $SO(N)$ Wilson loop in the antisymmetric representation given in Eq.~\eqref{eq:UxmoHnC0eB}. It is legitimate to directly adapt Yamaguchi's D5-brane from $AdS_5\times S^5$ to $AdS_5\times \RP^5$ for $\theta_k<\frac{\pi}{2}$ because, for these values of $\theta_k$, the $S^4$ lies in the upper hemisphere of $S^5$.\footnote{The equatorial case $\theta_k=\frac{\pi}{2}$, corresponding to $k=\frac{N}{2}$, is more subtle because antipodal points on the equator are identified under the quotient $S^5/\mathbb{Z}_2\cong \RP^5$. The fact that the volume of $\RP^4$ is half that of $S^4$ may indicate that the D5-brane dual to the antisymmetric Wilson loop corresponding to $k=N/2$ is doubly wrapped, or alternatively that the singly-wrapped D5-brane is dual to one of the two irreducible, self-dual representations into which the $k=N/2$ antisymmetric representation may be decomposed.} See Figure~\ref{subfig:2}.
\end{comment}

\section{The bremsstrahlung function for the spinor defect CFT}\label{sec:MOwof1jOHQ}

The 1/2-BPS Wilson loop defines an interesting example of a conformal defect labelled by a choice of representation of the gauge group. Using AdS/CFT, we can extract strong coupling predictions for the defect CFT$_1$ defined by the Wilson loop in the spinor representation by studying fluctuations of the dual D5-brane and their correlation functions. For the D3/D5-brane dual to the Wilson loops in the (anti)symmetric representations of $SU(N)$, this analysis was carried out in detail recently in \cite{Giombi:2020amn} both on the string and gauge theory sides. Our analysis in the present paper will be minimal: we restrict our attention to the two-point correlation functions of the displacement operators characterizing transverse displacements of the Wilson line, which are dual to transverse fluctuations of the D5-brane within $AdS_5$. For simplicity we will carry out the computation for the straight Wilson line, but we can easily map correlation functions on the straight Wilson line defect to correlation functions on the circular Wilson loop defect, when necessary. Our analysis of the fluctuating D5-brane is similar to the one in \cite{giombi2017half} for the  $AdS_2$ fundamental string. We leave a more detailed analysis along the lines of \cite{Giombi:2020amn} to future work. 

Defect correlation functions of operators $\mathcal{O}_1,\ldots,\mathcal{O}_n$ living on the 1/2-BPS spinor Wilson loop are defined to be:
\begin{align}
    \braket{\braket{\mathcal{O}_1(t_1)\ldots \mathcal{O}_n(t_n)}}:=\braket{{\rm Tr}_{sp} \mathcal{P}\left[\mathcal{O}_1(t_1)\ldots \mathcal{O}_n(t_n) e^{\int dt \left(iA_{\mu}\dot x^{\mu}+\Phi^6 |\dot x|\right)}\right]},
\end{align}
Certain operators have natural interpretations in terms of deformations of the contour $x^\mu(t)$ in $\mathbb{R}^4$ and of the scalar coupling $\Theta^I(t)$, defining the Wilson loop \eqref{eq:2W2phlg0Lb}. Here we focus on the displacement operators $\mathbb{D}_i(t)$ for the straight Wilson line, which generates transverse deformations of the curve in $\mathbb{R}^4$. They can be defined in terms of the variation of the Wilson loop in response to a variation in the $\mathbb{R}^4$ contour \cite{billo2016defects,correa2012exact}:
\begin{align}\label{eq:zsT1nHv3zl}
    \delta W&=\mathcal{P}\int dt \delta x^i(t) \mathbb{D}_i(t) W.
\end{align}
Here, $i=1,2,3$ denotes the three directions transverse to the unperturbed straight line contour $x^\mu(t)=(x^0(t),\ldots,x^3(t))=(t,0,0,0)$. By taking a variational derivative of Eq.~\eqref{eq:2W2phlg0Lb} with respect to $\delta x^j$, one finds $\mathbb{D}_i:= iF_{ti}+D_i\Phi^6$. Every defect CFT has displacement operators, whose scaling dimension is protected and equal to $\Delta_{\mathbb{D}}=2$ for a one-dimensional defect \cite{billo2016defects}. 
%the latter of which follows from Eq.~\eqref{eq:zsT1nHv3zl}.

On the string theory side, the fluctuations of the D5-brane dual to the spinor Wilson loop can be described by a field theory on $AdS_2\times \RP^4$. Turning off the fluctuations of the $U(1)$ gauge field (which is zero on the solution),\footnote{The gauge field fluctuations do not affect the leading order calculation of the displacement two-point function that we perform below, but they may affect four and higher point correlation functions, even if we restrict to the displacement sector for external states, as seen in the D5-brane calculations in \cite{Giombi:2020amn}.}  the bosonic part of the D5-brane action is given by the Nambu-Goto-like action
\begin{align}\label{eq:NG3JVSXB9S}
    S&=\frac{T_5L^6}{2}\int d^2 \sigma d^4 \tau \sqrt{{\rm det}_{\alpha\beta}\left[\frac{\partial_\alpha x^r \partial_\beta x^r+\partial_\alpha z\partial_\beta z}{z^2}+\frac{\partial_\alpha y^a\partial_\beta y^a}{\left(1+\frac{1}{4}y^2\right)^2}\right]},
\end{align}
where $\sigma^\alpha=(\sigma^1,\sigma^2,\tau^1,\ldots,\tau^4)$ are world-sheet coordinates on the brane, $x^r$, $r=0,\ldots,3$ are spacetime coordinates on the boundary of $AdS_5$ and $y^a$, $a=1,\ldots,5$, are stereographic spacetime coordinates %with $y^a\sim -y^a$ identified, 
on $\RP^5$. 
%Here, $h_{\alpha\beta}$ is the auxiliary worldsheet metric, which can be eliminated via its equation of motion in order to convert the Polyakov form into the Nambu-Goto form of the D-brane action.

Since here we will be interested only in the behavior of the displacement operators in the Wilson loop, which should be dual to the transverse fluctuations within $AdS_5$, we ignore the fluctuations of the D5-brane in $\RP^5$. Second, we pick coordinates on $AdS_5$ that make the $AdS_2$ geometry of the D5-brane manifest, as in \cite{giombi2017half}. 
The $AdS_5$ metric in such coordinates takes the form
%\begin{align}
%    ds^2&=ds^2_{AdS_5}+ds^2_{\RP^5},
%\end{align}
%where \cite{giombi2017half}
\begin{align}
ds^2_{AdS_5}&=\frac{\left(1+\frac{1}{4}x^2\right)}{\left(1-\frac{1}{4}x^2\right)^2}ds_2^2+\frac{dx^idx^i}{\left(1-\frac{1}{4}x^2\right)^2},&ds_2^2&=\frac{1}{z^2}\left(dx_0^2+dz^2\right)
\end{align}
%and the precise form of $ds_{\RP^5}^2$ is unimportant. 
Here $x^i$, $i=1,2,3$, are the coordinates characterizing transverse displacements of the $AdS_2$ part of the D5-brane. Finally, working in the static gauge (i.e., $\sigma^1=x^0$, $\sigma^2=z$, and the $\tau^1,\ldots,\tau^4$ are some parametrization of the static $\RP^4$ submanifold) and assuming the $x^i$ are constant on $\RP^4$ (i.e., considering only the lowest Kaluza-Klein modes in the dimensional reduction from $AdS_2\times \RP^4$ to $AdS_2$), we find that the effective Nambu-Goto action describing the transverse fluctuations of the D5-brane is:
\begin{align}
    S&=T_5L^6 V_{\RP_4}\int d^2 \sigma \sqrt{\text{det}_{\mu\nu}\left(\frac{\left(1+\frac{1}{4}x^2\right)}{\left(1-\frac{1}{4}x^2\right)^2}g_{\mu\nu}+\frac{\partial_\mu x^i\partial_\nu x^i}{\left(1-\frac{1}{4}x^2\right)^2}\right)},
\end{align}
where $\mu,\nu=1,2$ are the $AdS_2$ ``worldsheet" coordinate indices, $g_{\mu\nu}=\delta_{\mu\nu}/(\sigma^2)^2$, and $V_{\RP_4}=\frac{4}{3}\pi^2$ is the volume of $\RP^4$. We can expand the action in small $x^i$, and to quadratic order we find:
\begin{align}\label{eq:rISEKeyLSM}
    S&=T_5 L^6 V_{\RP^4}\int d^2\sigma \sqrt{g}\left(1+\frac{1}{2}\partial_\mu x^i\partial^\mu x^i+x^ix^i+O(x^4) \right).
\end{align}
Eq.~\eqref{eq:rISEKeyLSM} tells us that the $x^i$ fields are scalars in $AdS_2$ with mass $m^2=2$, corresponding to $\Delta=2$ using the familiar AdS/CFT dictionary $m^2=\Delta(\Delta-1)$. The $x^i$ are dual to the $\mathbb{D}_i$ displacement operators, which have the right scaling dimension, the same $SO(3)$ symmetry corresponding to rotations about the straight Wilson line, and the same interpretation in terms of transverse deformations of the Wilson line/surface.

Thus, by an $AdS_2/dCFT_1$ correspondence \cite{giombi2017half} between the defect CFT on the spinor Wilson loop and the fluctuations of the dual D5-brane, we expect:
\begin{align}\label{eq:rjxPw9e7nS}
\braket{\braket{\mathbb{D}_i(t_1)\mathbb{D}_j(t_2)}}_{CFT_1}=\braket{x^i(t_1)x^j(t_2)}_{AdS_2}.
\end{align}
We can use Eq.~\eqref{eq:rjxPw9e7nS} to find $\braket{\braket{\mathbb{D}_i(t_1)\mathbb{D}_j(t_2)}}_{CFT_1}$ at strong coupling. From the action in Eq.~\eqref{eq:rISEKeyLSM}, the boundary two-point function in $AdS_{d+1}$ can be readily calculated. In particular, the two-point function for a scalar field governed by the action $S=\frac{T}{2}\int d^d z dz_0 \sqrt{g}(g^{\mu\nu}\partial_\mu\phi \partial_\nu\phi+m^2\phi^2)$, with $m^2=\Delta(\Delta-d)$ and $g_{\mu\nu}=\frac{1}{z_0^2} \delta_{\mu\nu}$, is \cite{freedman1999correlation}:\footnote{Notice that the factor of $T$ appears in the numerator rather than the denominator of the two point function, the opposite of what one would find for 2-point functions in ordinary field theory. This follows from the definition of the $AdS$ correlation functions as variational derivatives of the exponential of the action with respect to the fields at the boundary.} 
\begin{comment}
Specifically, let $S[\phi;\phi\rvert_{\partial}=h]$ denote the quadratic $AdS_{d+1}$ action for the scalar field $\phi$, which approaches $z_0^{-\Delta} h$ near the boundary $\partial$ at $z_0=0$. Then, we have:
\begin{align}
    \frac{\delta }{\delta h(t_1)}\frac{\delta }{\delta h(t_2)}e^{-C S[\phi;\phi\rvert_{\partial}=h]}=\frac{\delta }{\delta h(t_1)}\frac{\delta }{\delta h(t_2)}e^{-S[\tilde{\phi};\tilde{\phi}\rvert_{\partial}=\sqrt{C}h]}=C\frac{\delta}{\delta \tilde{h}(t_1)}\frac{\delta}{\delta \tilde{h}(t_2)} e^{-S[\tilde{\phi};\tilde{\phi}\rvert_{\partial}=\tilde{h}]},
\end{align}
where $\tilde{\phi}=\sqrt{C}\phi$ and $\tilde{h}=\sqrt{C}h$.
\end{comment}
\begin{align}
\braket{O_\Delta(x)O_\Delta(y)}_{AdS_{d+1}}&=T\frac{(2\Delta-d)\Gamma(\Delta)}{\pi^{d/2}\Gamma(\Delta-\frac{d}{2})}\frac{1}{|x-y|^{2\Delta}}.
\end{align}
For the case of the $x^i$ fields, with $d=1$ and $\Delta=2$ and taking into account the factor of $T_5 V_{\RP^4}$ in the normalization of the action, this yields:
\begin{align}
    \braket{x^i(t_1)x^j(t_2)}&=\frac{6T_5L^6 V_{\RP_4}}{\pi}\frac{\delta_{ij}}{|t_1-t_2|^4}
=\frac{N\sqrt{\lambda}}{\sqrt{2}\pi^3}\frac{\delta_{ij}}{|t_1-t_2|^4}.\label{eq:SZorZvFXXE}
\end{align}

Because there is a canonical choice for the normalization of the $x^i$ fields according to Eq.~\eqref{eq:NG3JVSXB9S} and for the $\mathbb{D}_i(t)$ operators according to Eq.~\eqref{eq:zsT1nHv3zl}, the coefficient of the two-point function in Eq.~\eqref{eq:SZorZvFXXE} is a meaningful quantity that we can compare to physical properties of the boundary CFT. In particular, we can relate the normalization of the two-point function of the $x^i$ fields to the bremsstrahlung function of  the Wilson loop in the spinor representation, by using the results derived in \cite{correa2012exact}. Section $4$ of \cite{correa2012exact} argues for precise relationships between three observables in the defect conformal field theory associated with a Wilson line: the derivative of the anomalous dimension of a cusp in the Wilson line evaluated at zero angle, the normalization of the two-point function of the displacement operators, and the energy radiated by an accelerating quark.

Most relevant to our present purpose, the two-point function of the displacement operators $\mathbb{D}_i$ on a straight Wilson loop is related to the bremsstrahlung function, $B(\lambda)$, by:\footnote{The bremsstrahlung function is also related to the anomalous dimension of a cusp of angle $\phi$ by the equation $B=\frac{1}{2}\partial_\phi^2 \Gamma_{\text{cusp}}(\phi)\rvert_{\phi=0}$, and yields the energy of a moving quark moving at low velocities by $\Delta E=2\pi B\int dt (\dot{v})^2$.}
\begin{align}\label{eq:IQFtf0oyyQ}
    \braket{\braket{\mathbb{D}_i(t_1)\mathbb{D}_j(t_2)}}&=\frac{12 B(\lambda) \delta_{ij}}{t_{12}^4}.
\end{align}
From Eq.~\eqref{eq:rjxPw9e7nS}, Eq.~\eqref{eq:SZorZvFXXE}, and Eq.~\eqref{eq:IQFtf0oyyQ}, it follows that the leading contribution to the spinor Wilson loop bremsstrahlung function at strong coupling is:
\begin{align}\label{eq:Hi2EPEKX1J}
B_{sp}(\lambda)&=\frac{N\sqrt{\lambda}}{12\sqrt{2}\pi^3}+O\left(\frac{1}{\sqrt{\lambda}}\right).
\end{align}
%The $AdS_2\times \RP^4$ theory of fluctuating fields thus provides a simple derivation of the leading contribution to the bremsstrahlung function for the strongly coupled spinor Wilson loop.

We can also calculate the bremsstrahlung function directly in the defect CFT using supersymmetric localization. The argument is the same as for the case of the fundamental Wilson loop \cite{correa2012exact,giombi2017half}. First, one notes that the bremsstrahlung function can also be extracted from the two-point function of the scalar operators $\Phi^a$, $a=1,\ldots,5$, i.e. those that do not appear in the Wilson loop \eqref{eq:2W2phlg0Lb}. This is because these operators belong to the same supermultiplet of $OSp(4^*|4)$ as the displacement. Their two-point function is given by
\begin{align}
\braket{\braket{\Phi^a(t_1)\Phi^b(t_2)}}&=\frac{2B(\lambda)\delta_{ab}}{t_{12}^2}.
\label{2pt-Phi}
\end{align}
In particular, the dimensions of the $\Phi^a$ fields are protected and equal to $1$, and the normalizations of the two different two-point functions are related by a factor of $6$. These properties follow from supersymmetry and are independent of the gauge group and representation. %which is why we can transfer the results given in \cite{correa2012exact,giombi2017half} for the fundamental $SU(N)$ Wilson loop to the spinor $SO(N)$ Wilson loop. 

The normalization constant in (\ref{2pt-Phi}) can be extracted via localization by considering certain ``twisted" combinations of the scalars $\Phi^a$ that have position independent correlation functions on the circle, see \cite{correa2012exact, fiol2014exact, giombi2017half, giombi2018exact, liendo2018bootstrapping} for more details. The end result is that, for any representation $R$, one obtains the exact result 
\begin{equation}
B_R(\lambda) = \frac{\lambda}{2\pi^2} \frac{\partial}{\partial \lambda}\log \langle W_R\rangle \,.
\end{equation}

For the spinor Wilson loop, using Eq.~\eqref{eq:PWl70hirur}, this yields
\begin{align}
    B_{sp}(\lambda)&=\frac{\lambda}{2\pi^2} \frac{\partial}{\partial \lambda}
\frac{2N}{\pi}\int_{0}^1 \sqrt{1-u^2}\log\left(\cosh\left(\frac{1}{2}\sqrt{\frac{\lambda}{2}}u\right)\right)du \nonumber \\
%-\frac{\partial^2}{\partial A_1^2}\left(\frac{N}{\pi}\int_{-1}^1 %\sqrt{1-u^2}\log\left(\cosh\left(\frac{1}{2}\sqrt{\frac{\lambda'}{2}}u\right)\right)du\right)\biggr\rvert_{A_1=2\pi}\nonumber\\
&=\frac{N\sqrt{\lambda}}{4\sqrt{2}\pi^3}\int_{0}^1 \sqrt{1-u^2}u \tanh\left(\frac{1}{2}\sqrt{\frac{\lambda}{2}}u\right)du.
\label{Blam}
\end{align}
In the large $\lambda$ regime, this reduces to $B_{sp}(\lambda)=\frac{N\sqrt{\lambda}}{12\sqrt{2}\pi^3}+O\left(\frac{1}{\sqrt{\lambda}}\right)$, in precise agreement with Eq.~\eqref{eq:Hi2EPEKX1J}. 
Localization may be similarly applied to obtain predictions for higher-point functions of the protected scalar operators. It would be interesting to match them with the correlators obtained from the D5-brane. 

\section{Conclusions\label{sec:ewr5bkUoCk}}

In this paper, we studied the half-BPS circular Wilson loop in the ${\cal N}=4$ SYM theory with orthogonal gauge group. Using supersymmetric localization, we derived exact results for its expectation value, focusing on the large $N$ limit and on the fundamental, (anti)symmetric and spinor representations. In the latter case we obtained a simple quantitative test of the proposal of \cite{witten1998baryons} by matching the classical action of the $AdS_2\times \RP^4$ D5-brane with the strong coupling limit of the spinor Wilson loop expectation value. 

There are many natural extensions of the calculations we have discussed. One could try to study in detail the spectrum of fluctuations of the D5-brane dual to the spinor Wilson loop, along the lines of earlier work \cite{Harrison:2011fs, Faraggi:2011bb, Faraggi:2011ge} done for the D3/D5-branes dual to (anti)symmetric representations. Knowing the spectrum of excitations of the D5-brane would allow one to compute the one-loop correction to the Wilson loop expectation value, as in \cite{Faraggi:2014tna, Buchbinder:2014nia}, which would be very interesting to match with the localization prediction (for this, one would need to extend the analysis done in \S~\ref{sec:E4dda6KwCr} to order $N^0$). From the point of view of defect CFT, it would also be interesting to match the spectrum of excitations of the brane to the spectrum of defect operator insertions on the Wilson loop. Using localization and techniques similar to the ones developed recently in \cite{Giombi:2020amn}, one should be able to obtain exact results for a subsector of defect correlation functions and match them with the D5-brane dynamics. While the analysis should be somewhat similar to the one done in \cite{Giombi:2020amn} for the antisymmetric representation (dual to the $AdS_2\times S^4$ D5-brane), there should be important differences. For instance, in the spinor Wilson loop case there should be no ``Legendre transform" with respect to the gauge field, which is related in the case of (anti)symmetric representation to fixing the rank $k$ of the representation. It could also be worthwhile to see whether the defect spectrum and correlation functions on the spinor Wilson loop may be controlled by some underlying integrability. In the case of the antisymmetric Wilson loop in the $SU(N)$ theory, some steps along these lines were taken in \cite{Correa:2013em}. The case of the spinor representation should be similar, and perhaps simpler: in some sense, the spinor representation is more ``fundamental" than the rank-$k$ antisymmetric one because it does not arise from the tensor product of $k$ fundamental representations. 

Another extension natural from the defect CFT point of view would be to study ``bulk-defect" observables, i.e. correlation functions of the Wilson loop, with or without defect insertions, and local operators inserted away from the Wilson loop. Some such calculations were done in \cite{Giombi:2006de} for the (anti)symmetric representations of $SU(N)$, and it would be nice to extend them to the $SO(N)$ case and the spinor representation. The ``bulk-defect" correlation functions play an important role in the bootstrap approach \cite{liendo2018bootstrapping} to defect CFT, and are also accessible by supersymmetric localization when restricted to a special protected subsector.   

Finally, it may also be interesting to study other representations of the orthogonal gauge group, like the spin-$s$ tensor-spinor representations $[s,1/2,\ldots, 1/2]$ (with $s$ a half-integer), and see whether they have a natural D-brane description on the string theory side. The analysis of larger representations of size $\sim N^2$ (that may also include large spinorial representations), would also be of interest: they should be dual to ``bubbling geometries" with $AdS_5\times \RP^5$ asymptotics, similarly to the $AdS_5\times S^5$ case studied in \cite{Yamaguchi:2006te, Lunin:2006xr, DHoker:2007mci},\footnote{The bubbling geometries dual to large chiral primary local operators in the case of $SO(N)$/$USp(N)$ gauge theory was studied in \cite{Mukhi:2005cv} and further discussed in \cite{fiol2014exact}.} but they may involve new features that are worth investigating.

\section*{Acknowledgments}

We thank Shota Komatsu for useful discussions and collaboration on related topics. This work was supported in part by the US NSF under Grants No.~PHY-1620542 and PHY-1914860. 

\appendix
\section{Quadratic Casimirs of $SO(N)$}
The quadratic Casimir of $SO(N)$ appears in the leading term in the Feynman diagram expansion of the Wilson loop, given in Eq.~\eqref{MMpert}. Presently, we collect the expressions for the quadratic Casimirs in the fundamental, the symmetric, the antisymmetric, and the spinor representations.

As in the main text, let $T^a_R$ be the generators of $SO(N)$ in the representation $R$ of dimension $\text{dim}(R)$. The quadratic Casimir is defined by $T^a_RT^a_R=C_2(R)1_{\text{dim}(R)\times \text{dim}(R)}$.
% We define the normalization of the generators, $C(R)$, and the quadratic Casimir, $C_2(R)$, via the relations $\text{Tr}_R(T^a_RT^b_R)=C(R)\delta^{ab}$ and $T^a_RT^a_R=C_2(R)1_{d(R)\times d(R)}$. 
Once we fix the normalization of the generators in the fundamental representation by $\text{Tr}(T^a_F T^b_F)=C(F)\delta^{ab}$, $C_2(R)$ is fixed in terms of $C(F)$, the Dynkin labels $a_i$ of $R$ and the inverse Cartan matrix $G_{ij}$ by the equation (see Eq.~27 in \cite{van1999group}):
\begin{align}\label{eq:pIRw00LbIw}
    C_2(R)=\frac{C(F)}{2}\sum_{i=1}^{N'} \sum_{j=1}^{N'} (a_i+2)G_{ij}a_j.
\end{align}
Here, $N'$ is the rank of $SO(N)$ when $N=2N'$ or $N=2N'+1$.

To determine $C_2(R)$ for the four representations, we need the inverse Cartan matrix for $SO(N)$ and the Dynkin labels for each representation. The inverse Cartan matrices for $SO(2N')$ and $SO(2N'+1)$ are given in Table~3 of \cite{Yamatsu:2015npn}. Furthermore, the Dynkin labels generically can be related to the Young tableau indices $m_i$, $i=1,2,\ldots,N'$, which specify the number of boxes in the $i$th row of the representation's Young tableau. Section~1.58 of \cite{frappat2000dictionary} tells us that when $N=2N'$, then $a_i=m_i-m_{i+1}$ for $1\leq i<N'$ and $a_{N'}=m_{N'-1}+m_{N'}$; and when $N=2N'+1$, then $a_i=m_i-m_{i+1}$ for $1\leq i<N'$ and $a_{N'}=2m_{N'}$. Using the explicit forms of $G_{ij}$ and the relations between the $a_i$ and $m_i$ for both even and odd $N$, we can rewrite Eq.~\eqref{eq:pIRw00LbIw} in a more useable form:
\begin{equation}\label{eq:L9U2AyKsnw}
C_2(R) = \frac{C(F)}{2}\left(N r + \sum_{i=1}^{N'} m_i (m_i-2i) \right),
\end{equation}
where $r=\sum_i m_i$. 

Finally, we consider the four specific representations. For the fundamental representation, $m_1=1$, $m_i=0$ for $2\leq i\leq N'$. For the rank $k$ symmetric representation, $m_1=k$, $m_i=0$ for $2\leq i\leq N'$. For the rank $k$ antisymmetric representation, $m_i=1$ for $1\leq i\leq k$ and $m_i=0$ for $k<i\leq N'$. Finally, for the spinor representation, $m_i=1/2$ for $1\leq i\leq N'$. It follows that the non-zero Dynkin labels are $a_1=1$ for the fundamental, $a_1=k$ for the symmetric, $a_k=1$ for the antisymmetric, and $a_{N'}=1$ for the spinor. Using either Eq.~\eqref{eq:pIRw00LbIw} or Eq.~\eqref{eq:L9U2AyKsnw}, we arrive at the following expressions:
\begin{align}
    C_2(F)&=C(F)\frac{N-1}{2},\\
    C_2(S_k)&=C(F)\frac{k(k+N-2)}{2},\\
    C_2(A_k)&=C(F)\frac{(N-k)k}{2},\\
    C_2(sp)&=C(F)\frac{N(N-1)}{16}.
\end{align}


\bibliographystyle{ssg}
\bibliography{mybib}

\end{document}